\PassOptionsToPackage{pdfpagelabels=false}{hyperref} 

\documentclass[twocolumn]{aastex631}
\usepackage{amsmath}
\usepackage{amssymb}
\usepackage{xcolor}

\shorttitle{Bounce to Breakout}
\begin{document}
\title{}

\title{A 3D Simulation of a Type II-P Supernova: from Core Bounce to Beyond Shock Breakout} 
\correspondingauthor{David Vartanyan}
\email{dvartanyan@carnegiescience.edu}

\author[0000-0003-1938-9282]{David Vartanyan}
\affiliation{Carnegie Observatories, 813 Santa Barbara St., Pasadena, CA 91101, USA; NASA Hubble Fellow}
\author[0000-0002-6543-2993]{Benny T.-H. Tsang}
\affiliation{Department of Astronomy and Theoretical Astrophysics Center, University of California, Berkeley, CA 94720, USA}
\author[0000-0002-5981-1022]{Daniel Kasen}
\affiliation{Department of Astronomy and Theoretical Astrophysics Center, University of California, Berkeley, CA 94720, USA}
\author[0000-0002-3099-5024]{Adam Burrows}
\affiliation{Department of Astrophysical Sciences, Princeton University, NJ 08544, USA; School of Natural Sciences, Institute for Advanced Study, Princeton, NJ 08540}
\author[0000-0002-0042-9873]{Tianshu Wang}
\affiliation{ Department of Physics, University of California Berkeley, Berkeley, CA, 94720}
\author[0009-0009-3068-9527]{Lizzy Teryoshin}
\affiliation{University of California, San Diego, CA, 92037, USA}



\begin{abstract}
In order to better connect core-collapse supernova (CCSN) theory with its observational signatures, we have developed a simulation pipeline from the onset of core collapse to beyond shock breakout from the stellar envelope. Using this framework, we present a three-dimensional simulation study from five seconds to over five days following the evolution of a 17-M$_{\odot}$ progenitor, exploding with $\sim$10$^{51}$ erg of energy and $\sim$0.1 M$_{\odot}$ of $^{56}$Ni ejecta. The early explosion is highly asymmetric, expanding most prominently along the southern hemisphere. This early asymmetry is preserved to shock breakout, $\sim$1 day later. Breakout itself evinces strong angle-dependence, with as much a day delay in shock breakout by direction. The nickel ejecta closely tail the forward shock, with velocities at breakout as high as $\sim$7000 km\,s$^{-1}$. A delayed reverse shock forming at the H/He interface on hour timescales leads to the formation of Rayleigh-Taylor instabilities, fast-moving nickel bullets, and almost complete mixing of the metal core into the hydrogen envelope. For the first time, we illustrate the angle-dependent emergent broadband and bolometric light curves from simulations evolved in 3D in entirety, continuing through hydrodynamic shock breakout a CCSN model of a massive stellar progenitor evolved with detailed, late-time neutrino microphysics and transport. Our case study of a single progenitor underscores that 3D simulations generically produce the cornucopia of observed asymmetries and features in CCSNe observations, while establishing the methodology to study this problem in breadth.
\end{abstract}


\section{Introduction}
\label{sec:int}
Core-collapse supernova (CCSN) theory lies at a crossroads. More than half a century after the earliest CCSN simulations \citep{1966ApJ...143..626C}, and enabled by improvements in stellar evolution, microphysics, and computational capability, current simulations are able to produce successful explosions driven by the neutrino heating mechanism \citep{1985ApJ...295...14B} with early forays into observational predictions \citep{janka2012,2021Natur.589...29B}. Emboldened by heady progress, CCSN theory is ready to confront with modeling capabilities all-sky searches for supernovae and their remnants.

Supernova structures have long been known to be multi-dimensional, as evidenced by spectropolarimetric signatures  \citep{2008ARA&A..46..433W,2006Natur.440..505L},
aspherical nickel distributions in supernovae remnants revealed by light echoes \citep{sinnott,rest} and fine structure in line profiles \citep{1988MNRAS.234P..41H}, and direct remnant imaging \citep{2024ApJ...965L..27M,2019ApJ...886..147L}.


3D mapping of supernovae remnant Cassiopeia A (Cas A) reveals a prominent, spherical reverse shock with several dominant plumes, driven by nickel expansion, which curl into `fingers'. Rings in the exterior shock ejecta wrap around several large expanding nickel bubbles, and the interior, unshocked ejecta is dotted with high-velocity `knots' within cavities \citep{2002A&A...381.1039W, delaney,2012ApJ...751...25M,2014Natur.506..339G}. Imaging of radioactive titanium in the interior unveiled structure in the innermost ejecta, yet untouched by the reverse shock.




Multi-band observations of SN1987A similarly reveal inherent multi-scale asymmetries beyond its famous three-ring structure. These observations span radio \citep{2010ApJ...710.1515Z}, probing the shock interaction with the circumstellar environment; sub-millimeter, highlighting the distribution of freshly-synthesized dust \citep{2012A&A...541L...1L}; infrared \citep{2010ApJ...722..425D}, illustrating the gas-grain interactions along the SN shock; optical, depicting the ejecta distribution by elements and the shock interaction with the nebular rings, dotted with ``hot spots"; and X-ray to gamma-ray  \citep{2015Sci...348..670B,2011Natur.474..484L,1988Natur.331..416M}, powered by the radioactive decay of $^{56}$Ni, $^{56}$Co, and $^{44}$Ti. Early X-ray and gamma-ray observations indicate significant $^{56}$Ni mixing \citep{1988Natur.332..516L}. Outwardly mixed, rapidly moving $^{56}$Ni bullets are often invoked to explain details in H$\alpha$ spectroscopy in the famous `Bochum' event \citep{1988MNRAS.234P..41H, 2002ApJ...579..671W} in the first month after breakout.

Recent redoubled efforts with high-resolution infrared imaging via the \textit{James Webb Space Telescope} provided new perspectives into the cornucopia of asymmetries present in Cas A \citep{2024ApJ...965L..27M} and SN1987A \citep{2024MNRAS.532.3625M}. Observation of central emission lines from SN1987A could be associated with a kicked cooling neutron star or a pulsar wind nebula \citep{2024Sci...383..898F}, providing constraints on its unresolved compact object. 

Keeping pace with observational developments, multiple groups using 3D CCSN simulations with different codes and methodologies have recently succeeded in producing explosions via the neutrino heating mechanism \citep{vartanyan2018b,2021Natur.589...29B,burrows_2020,burrows2024,2021ApJ...915...28B, muller2017,glas2019,roberts:16,ott2018_rel,lentz:15}. Advances in simulation capabilities come jointly with improvements in CCSN theory, particularly in neutrino microphysics, and the role of the stellar progenitor interior structure, particularly the silicon-oxygen interface, in prompting explosion \citep{burrows2018,2021ApJ...916L...5V,vartanyan2018a,2024Univ...10..148B,janka2012,tsang2022,wang,2023ApJ...949...17B}.

Reconciling CCSN theory with observations mandates a multi-physics, multi-scale (spatial and temporal) strategy that transcends kilometer- and microsecond scales in the stellar core to solar radii through parsec scales (e.g., the spatial extent of Cas A, produced with a forward shock velocity of $\sim$5800 km s$^{-1}$ enduring for three centuries, \citealt{vink}), coupling the nuclear and neutrino physics of the collapsing core with the hydrodynamic evolution of the nucleosynthesized material. CCSNe are multi-messengers, with light-curves and panchromatic electromagnetic, neutrino \citep{2023MNRAS.526.5900V}, and gravitational wave \citep{2023PhRvD.107j3015V,choi2024} signatures. 

The earliest simulation studies of supernovae to breakout \citep{1973MNRAS.161...47G}, including 2D simulations \citep{1991ApJ...367..619F,1989ApJ...341L..63A}, identified the development of hydrodynamic instabilities \citep{1992ApJ...392..118C} at composition shell interfaces and mediated by the reverse shock, but did not consistently model the development of the explosion and the dominant neutrino-driven instabilities, which seed the early CCSN asymmetries. 
Subsequent 2D simulations of CCSNe out to late times date to  \citet{kifonidis2000, kifonidis2003}, using a parametrized neutrino bomb to prompt explosion, illustrated that nickel clumps decoupled from the shock and moved ballistically through the stellar envelope with velocities up to several 1000 km\,s$^{-1}$. Later studies were also similarly parametrized in 2D \citep{joggerst} and in 3D \citep{hammer} with gray neutrino transport. 3D simulations are necessary to capture the efficient mixing of metal-core elements into the stellar envelope, induce the Rayleigh-Taylor instability (RTI) at physical rates, capture explosion asymmetry, and accelerate ejecta to observed high velocities. The RTI is triggered as a hydrodynamic instability across density gradients, often composition interfaces (hydrogen/helium, H/He, in our study) and associated with shock deceleration as the shock sweeps up a massive envelope.

Collectively, early simulations excised the proto-neutron star (PNS) core during the crucial earlier stages of CCSN evolution. PNS convection contributes significantly to the neutrino luminosity and to the inner turbulent hydrodynamics in the first seconds of CCSN evolution \citep{2006ApJ...645..534D,radice2017b,nagakura_pns} and necessitates self-consistent modeling.   

3D studies of shock breakout grew in abundance in the last decade, with varying levels of detail in the CCSN simulation, especially in the treatment of the core and the sophistication of neutrino-matter coupling.
\citet{2015ApJ...810..168O} and \citet{wongwathanarat2015} mark a transition from artificial explosion triggers towards a more self-consistent treatment of neutrino physics and transport, although still relying on imposed neutrino luminosities.
 Using Prometheus-HOTB with an approximate gray, ray-by-ray scheme for neutrino transport on an axis-free `Yin-Yang' grid, \cite{wongwathanarat2017} studied
 in 3D the evolution of  a15-M$_{\odot}$ Cas A-like CCSN through breakout, later continued for centuries \citep{orlando-cent} and millennia \citep{orlando-mil}. The ray-by-ray approach neglects lateral transport of neutrinos, rather solving along many `1D' rays, and may artificially promote explosion in 2D simulations \citep{dolence_2015,skinner2016,vartanyan2018a} and perhaps in 3D as well \citep{glas2018}. Neutrino-driven convection, which drives the initial asymmetry in explosion, was seeded by 0.1$\%$ amplitude random radial velocity perturbations. Late-time models, including MHD evolution with the PLUTO code to study supernova remnant (SNR) interaction with the circumstellar material (CSM) and interstellar material (ISM), enable direct insight via observational templates into pre-collapse stellar evolution, progenitor properties, CCSN mechanism, and the emergent diagnostics \citep{2024arXiv240812462O}.

Again using approximate gray ray-by-ray neutrino transport, \cite{gabler} evolved several red and blue supergiant progenitors (RSG, BSG) of 15 and 20 M$_{\odot}$ \citep{wongwathanarat2013} out to one year. They found iron-ejecta velocities accelerating by a few hundred km\,s$^{-1}$ to 1000 km\,s$^{-1}$, driven by radioactive decay of the nickel chain and a rebounding reverse shock, and emphasize a strong correlation between both the shock structure and the nickel ejecta at shock revival, on sub-second timescales, and at one year. 

\cite{utrobin} modeled a series of BSG progenitors using the same setup as \cite{wongwathanarat2013,wongwathanarat2015} summarized above, with explosion imposed artificially by introducing a neutrino luminosity (and energy) boundary condition in the stellar core. Using a series of binary-merger progenitor models, the authors were able to reproduce 11 out of their selection of 12 quantified observational constraints of SN1987A, matching properties of both the progenitor star Sanduleak -69$^{\circ}$202 and its explosion. However, whether self-consistent models can produce similar successful results in explaining observed properties of CCSNe remained unanswered.

Breakout studies explored red supergiants (RSG), blue supergiants (BSG), and more recently, helium cores, in all cases highlighting the necessity of multidimensionality to robustly explain emergent properties. \cite{2018MNRAS.479.3675M} evolved until breakout a binary ultra-stripped 2.8-M${_\odot}$ helium star through CCSN explosion with Coconut-FMT, a spherical-polar general-relativistic code  with simplified multi-group neutrino transport, finding small energies, low kick velocities, and significant mixing, which is inadequately explained by a mixing-length treatment.

Several studies continued shock breakout simulations to follow black hole formation. \citet{2018ApJ...852L..19C,moriya} study a zero-metallicity 40\,M$_{\odot}$ progenitor evolved with Coconut-FMT and continued until shock breakout with the moving-mesh hydrodynamic code AREPO. They found joint formation of a weak explosion and a black hole.  \cite{chan} repeat the study for a 12-M$_{\odot}$ progenitor and a higher-energy 40-M$_{\odot}$ model.
\cite{burrows2023} evolved a solar-metallicity 40-M$_{\odot}$ progenitor and similarly saw joint black hole formation with shock revival, but with an energetic $\sim$1.6 B explosion. 
\cite{rahman} evolve three pulsational pair-instability supernovae progenitors with masses of 60, 80, and 115 M$_{\odot}$ using their general-relativistic flux-limited diffusion code NADA-FLD for core collapse and Prometheus to study subsequent evolution. All models except their rapidly-rotating 60-M$_{\odot}$ model experienced shock revival. Collapse to a black hole  proceeded shortly afterwards, but with a weaker sustained neutrino emission for several hundred milliseconds. All neutrino-heated material is accreted, but a shock or weakly-resolved sonic pulse may still propagate outward and eject mass. Even in failed CCSNe, some mass loss is expected \citep{1980Ap&SS..69..115N,2018MNRAS.477.1225C}. In a recent breakout study performed in 2D-axisymmetry, \citet{2024arXiv241004944S} explored the fallback masses for very massive progenitors (60$-$95\,M$_{\odot}$), which also exploded and formed black holes.

During the last several years, breakout studies with more detailed neutrino heating have emerged. \cite{stockinger} studied low-mass, Crab-like progenitors by coupling Prometheus-Vertex for CCSN evolution with Prometheus-HOTB, evolving through shock breakout. Unlike earlier models, these models include detailed neutrino transport in 3D, but maintain the ray-by-ray approximation. Due to the computational expense, neutrino transport is approximated by a simplified heating/cooling lightbulb-like scheme (NEMESIS) after 0.5 seconds. This approximation may not be too severe for low-mass models \citep{burrows_2019}, which generally saturate to low explosion energies early on.  The authors find that an extended hydrogen envelope allows for more efficient mixing and larger-scale asymmetries as nickel plumes deform the shock front.

More recently, \cite{sandoval} performed a series of 2D and 3D simulations carrying out CCSN models of two low-mass progenitors, a zero-metallicity 9.6-M$_{\odot}$ progenitor and a 10-M$_{\odot}$ progenitor, evolved with CHIMERA and continued through shock breakout with FLASH. CHIMERA evolves detailed neutrino transport in 3D, but also with the ray-by-ray approximation. These simulations boast both a higher angular and radial resolution and a larger nuclear network (160 species) than many earlier studies, and conclude explosion morphology is strongly influenced by the dynamics of metal-rich ejecta.

The CCSN problem can be stated as follows: interpreting the zoo of CCSN observation requires disentangling the core explosion and its morphology from subsequent interaction and evolution with the stellar envelope, the CSM, and the ISM. A key theme from the decades of breakout studies is that large hydrodynamic instabilities grow from small ones seeded at shock
revival via neutrino-driven convective turbulence. At breakout, how much of the observed asymmetry is caused by the CCSN itself, and how much is due to its environment?  A quantitative answer requires an integrated multi-scale effort. Pressingly, 3D simulations with detailed neutrino heating to late times, as explosion energies begin to asymptote, have been entirely lacking. Long-term CCSN simulations are required to produce both robust final explosion energies and nucleosynthetic compositions, both of which require many seconds to reach their final values \citep{wang2024, burrows2024,muller2017}.  Particularly, late-time 3D CCSN simulations of massive progenitor with $\sim$Bethe ($\equiv$10$^{51}$ ergs) explosion energies and detailed neutrino microphysics and transport are necessary, but entirely absent. 


Despite the progress in CCSN simulations out to shock breakout, several significant simplifications persist.  As illustrated, a limited set of breakout simulations exists. Few models amongst this set are multi-dimensional, and fewer still are explosions driven with self-consistent neutrino heating. Detailed breakout simulations with predictive yields, including resulting photometry and spectra, are exceedingly rare. In this paper, we present such a study, one of the first 3D explosion-to-breakout simulations, starting from late-time modeling of the explosion with detailed neutrino heating, and continued out out to nearly homologous expansion with long-term light-curves and spectra. The small-scale neutrino physics details at early seconds can determine the large-scale ejecta structure at days. Thus, a proper understanding of CCSN observations requires carefully modeling both in tandem. Our intent here is to make direct connections to observations for a single model, while setting the stage to perform a population study using a suite of models.

We present here a study of the energetic and asymmetric explosion of a 17-M$_{\odot}$ model, chosen as a 1 Bethe explosion within a red supergiant progenitor yielding a Type II-P supernova, characterized by a lengthy plateau phase due to an extended shock-ionized hydrogen envelope post-breakout. Large asymmetries are generally associated with high-energy explosions \citep{burrows2024}. We couple a detailed, late-time 3D neutrino radiation-hydrodynamic CCSN simulation using F{\sc{ornax}} and continued to follow shock propagation in the outer envelope beyond shock breakout using FLASH in order to provide post-breakout diagnostics and early predictive light signatures. We introduce our strategy and methods in \S\,\ref{sec:methods} and in \S\,\ref{sec:results} discuss our results, including the forward and reverse shock dynamics and the elemental distributions. Our study extends into the CSM in \S\,\ref{sec:csm} to provide early observational signatures in \S\,\ref{sec:sedona}. Finally, we summarize with our key findings in \S\,\ref{sec:conclusions}.

\section{Methods}\label{sec:methods}
\begin{figure*}
\centering
    \includegraphics[width=0.8\textwidth]{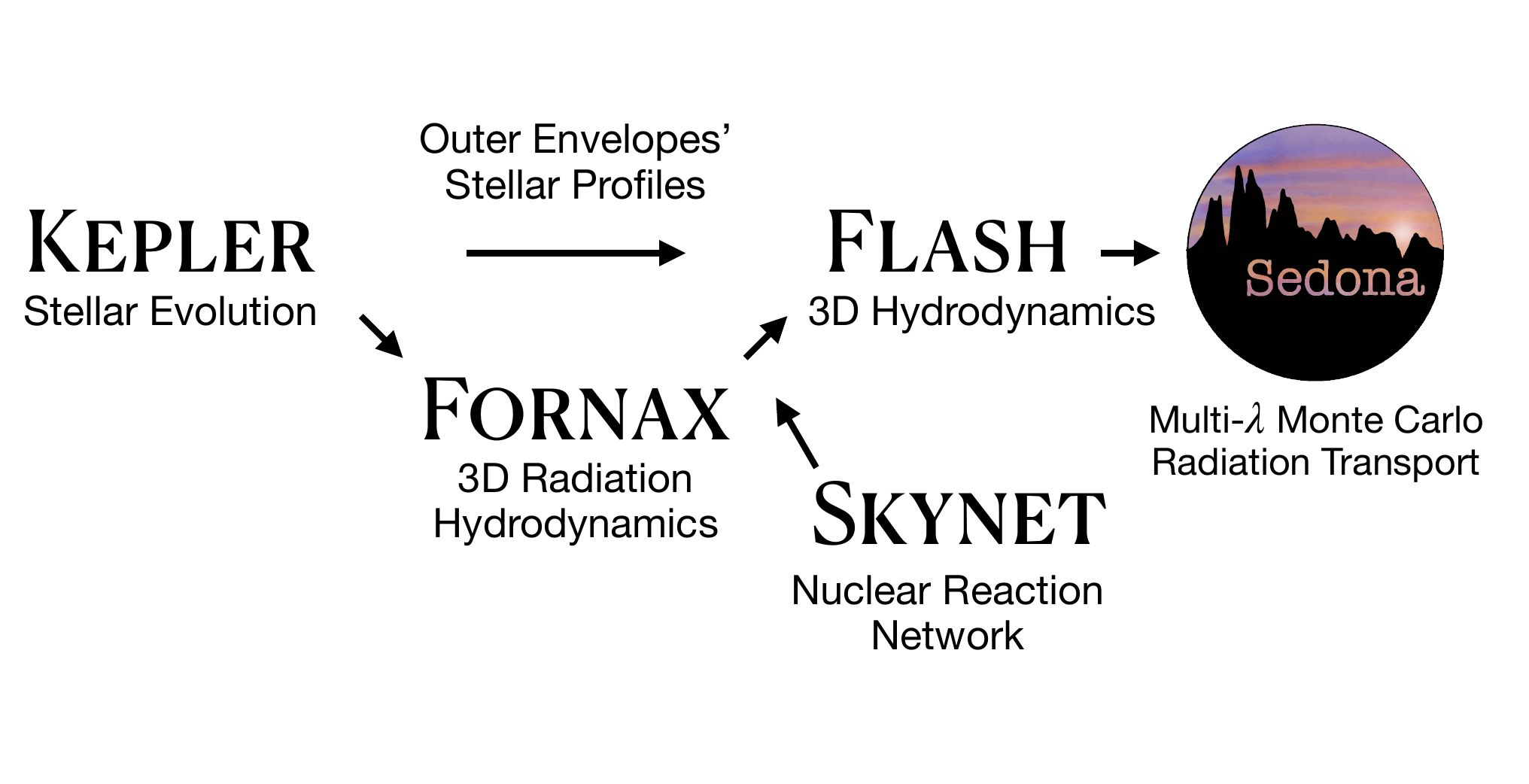}    
    \vspace{-1cm}
   \caption{ An illustration of our workflow. We take stellar evolutionary progenitors, currently using the KEPLER code, and follow their radiation-hydrodynamic evolution as CCSNe with F{\sc{ornax}} until neutrino heating subsides and the explosion energy asymptotes. We then add tracers in post-processing to calculate nuclear yields via SKYNET, and map to FLASH where we continue the simulation until breakout into the CSM. Finally, we predict spectral templates, line profiles, and light-curves with Sedona.}
    \label{fig:workflow}
\end{figure*}

We have built a pipeline to study stellar evolution from core collapse to beyond shock breakout, illustrated in Fig.\,\ref{fig:workflow} with key properties summarized in Table\,\ref{tab:sn_props}. 
We present a study of a 17-M$_{\odot}$ progenitor, which was evolved in 3D through 5.56 seconds post-bounce (and continuing, \citealt{burrows2024,2023PhRvD.107j3015V}) using the radiation-hydrodynamic supernova code F{\sc{ornax}} \citep{skinner2019}. The initial conditions were taken from the KEPLER stellar evolutionary code (\citealt{sukhbold2018}), with an infall velocity of $\sim$1000 km s$^{-1}$ identifying the onset of core collapse. The model was evolved in F{\sc{ornax}} on a grid extending out to 100,000 km (with an innermost cell size of 0.25 km) with 1024 $\times$ 128 $\times$ 256 cells in $r$, $\theta$, $\phi$ with multi-group M1 closure for the radiation transport and a detailed suite of neutrino microphysics, including the many-body correction to neutrino-nucleon inelastic scattering rate and neutrino scattering off both electrons and nucleons. At the end of the simulation, the 17-M$_{\odot}$ model had a shock radius spanning 30,000$-$76,000 km, and an explosion energy of $\sim$1.1 B. We choose this particular progenitor because of its explosion energy, nickel yield, and mass $-$ the model skirts the putative `red-supergiant problem' (\citealt{2009MNRAS.395.1409S}, but also \citealt{2011MNRAS.412.1522S,2018MNRAS.474.2116D}), the observational apparent dearth of CCSNe with RSG progenitor masses greater than $\sim17-18$ M$_{\odot}$.

We inject $\approx$320,000 post-processed tracers at the end of the F{\sc{ornax}} simulation. The tracers are placed logarithmically along the r-direction above 500 km and uniformly along the $\theta$- and $\phi$-directions and evolved backwardly \citep{sieverding2023b} with an adaptive sub-iteration method \citep{wang2023}. The tracer trajectories are then fed to SkyNet \citep{lippuner2017}, and a 1530-isotope network including elements up to  A = 100 is used to calculate the nucleosynthesis results. The reaction rates are taken from the JINA Reaclib \citep{cyburt2010} database, and we include neutrino interactions with protons and neutrons, but not reactions for the $\nu$-process \citep{woosley1990}. The nuclear statistical equilibrium (NSE) criterion is set at 0.6 MeV ($\sim$7 GK), and SkyNet switches to its NSE evolution mode if the temperature is above this mark and the strong-interaction timescale is shorter than the timescale of density changes. The electron fractions (Y$_e$) are calculated by F{\sc{ornax}} when the temperature is above the NSE threshold, which allows the neutrino spectra to be appropriately non-thermal. The nucleosynthesis calculation starts from the point after which the tracers never reach NSE again.
The initial abundances of a tracer are determined by the NSE if its temperature has ever reached the NSE criterion, otherwise the initial abundances are taken from the progenitor models \citep{swbj16,sukhbold2018}.

The 17-M$_{\odot}$ model is mapped to and continued with the general-purpose multi-physics code FLASH to follow its hydrodynamic evolution through shock breakout and into the CSM. FLASH \citep{2000ApJS..131..273F,2009arXiv0903.4875D} has seen use  in shock breakout calculations (e.g., \citealt{sandoval,2020ApJ...888..111O,joggerst}, with parametrized explosions in the latter two cases). We build on the FLASH methodology introduced in \cite{sandoval}, using the Split PPM (Piecewise-parabolic Method) solver with an HLLE Riemann solver. The simulation is run in spherical geometry with a resolution of 2240$\times$192$\times$384 in $r$, $\theta$, and $\phi$, with logarithmic spacing in radius. We select 22 isotopes with the highest abundances, including $\alpha$- and nickel-group elements and follow their advection in FLASH, renormalizing their summed mass fractions to one. The isotopes and physical variables are interpolated from the F{\sc{ornax}} to the FLASH grid. Our initial inner boundary in FLASH is 500 km, and the outer boundary is $\approx$7.03$\times10^{8}$ km, giving a radial resolution $\Delta{r}$/r $\sim$ 6.3$\times10^{-3}$. The angular resolutions are better than one degree. Our radial and angular resolutions compare favorably with models in literature (e.g. \citealt{sandoval} $\sim$ 6$\times10^{-3}$, $<$1$^{\circ}$; \citealt{stockinger}  $\sim$ 9$\times10^{-3}$, 2$^{\circ}$; \citealt{wongwathanarat2015} $\sim$ 1$\times10^{-2}$, 2$^{\circ}$; \citealt{2018MNRAS.479.3675M}  $\sim$ 9$\times10^{-3}$, 1.6$^{\circ}$). 

We impose a diode and outflow inner and outer radial  boundary condition, respectively, with periodic azimuthal and reflective polar boundaries.  To avoid the restrictive Courant condition along the poles, we excise a half-opening angle of 5$^{\circ}$. Because the F{\sc{ornax}} simulation does not include the outer envelope, we stitch on data from the KEPLER progenitor exterior to 98,000 km. In addition, we periodically excise cells from the inner grid to keep the inner boundary at $\sim$1$-$2$\%$ of the minimum shock radius to accelerate the simulation. We found that varying the frequency of excision and the inner boundary radius to have a negligible effect on the supernova evolution. Matter in the excised cells is bound and infalling and the contribution to the point mass, though small, is accounted for. 

The gravitational effects of the matter inward of the evolving inner boundary km are represented by a point mass, and Newtonian self-gravity is calculated with a multipole solver. Mapping is performed with good conservation of mass and energy.  We use an inflow boundary and do not account for any outgoing neutrino-driven wind (but see \citealt{2024arXiv241207831B} for models we evolved in FLASH with a wind to study fallback accretion). For this model, we find $\sim$0.7 M$_{\odot}$ of fallback accretion, almost entirely in the first hours, to a final remnant mass of $\sim$2.8 M$_{\odot}$. Although the neutrino-driven simulation initializing our FLASH follow-up is carried out to past five seconds, later than any other competing 3D CCSN efforts, a neutrino-driven wind at the inner boundary may still affect the early-time fallback accretion post-mapping \citep{wongwathanarat2015,2022ApJ...926....9J}. 

The F{\sc{ornax}} simulation uses the SFHo equation of state \citep{2013ApJ...774...17S}, generally consistent with most known theoretical, laboratory, and observationally-motivated nuclear physics constraints \citep{2017ApJ...848..105T, 2023Parti...6...30L}. The high-density, high-temperature inner core is excised, and we smoothly continue the simulation in FLASH with the Helmholtz equation of state \citep{2000ApJS..126..501T}, which includes internal energy contributions from ions, electrons, positrons, and radiation. The 22 isotopes studied are coupled into the Helmholtz equation of state.

Subsequently, we follow the evolution post-breakout in a two-fold manner. We continue our simulation on the original grid to follow the reverse shock propagation and ongoing mixing on several-day timescales. Secondly, we stitch on a toy CSM profile to the grid to both track the hydrodynamic evolution post-breakout, as the ejecta approach homologous expansion, and to describe the observed properties of the emergent electromagnetic signatures.

To prepare for radiation transfer post-processing, we remapped Flash model at the time $t_0 \equiv 264,000$ seconds onto a regular 50x50x50 Cartesian grid. At this time, the ejecta velocity structure is nearly linear, $\vec{v}(r) \propto \vec{r}$, except in the innermost layers ($r \lesssim 0.2 r_{\rm out} $) where a reverse shock persists. The thermal energy of the ejecta at this time is $3.3 \times 10^{50}$~erg, or about $\sim 50\%$ of the kinetic energy of $7.0 \times 10^{50}$~erg, indicating that some additional acceleration of the ejecta will occur before the system reaches true homologous expansion.

To avoid having to carry out a 3D coupled radiation-hydrodynamics calculation, we began our calculation at $t_{\rm exp} = 20$~days when the flow should be freely expanding and homologous. We applied an approximate technique to evolve 
the system from $t_0$ to $t_{\rm exp}$. Because radiation diffusion in the bulk of ejecta should be minimal in the first 20 days, the expansion should be nearly adiabatic; the thermal energy is radiation dominated and so will decline with radius as $1/r$ and so as $1/t$ for a homologous flow. We thus reduced the thermal energy content by a factor of $t_{\rm 0}/t_{\rm exp}$, giving a value of $4.5 \times 10^{49}$~erg.  The kinetic energy of the ejecta increased by a corresponding amount to account for the acceleration from $pdV$ work.  We then enforced a strictly homologous velocity structure, $\vec{v}(r) = \vec{r}/t_{\rm exp}$. 

The homologous ejecta structure was then input into the Sedona code \citep{sedona}, a Monte Carlo radiation transport tool. 
The initial radiation field was represented by discrete photon packets which were randomly distributed throughout the ejecta according to the initial thermal energy content of the model, and assuming that the radiation field was everywhere a blackbody at the local  temperature. At each subsequent time step, photon packets were emitted based on the radioactive energy release of $^{56}$Ni. The radioactive packets were sampled to be either gamma-rays or positrons, depending on the probability of emission of either type of particle. Positrons were assumed to deposit their energy immediately, whereas the gamma-ray packets were transported until their energy was lost due to Compton scattering of photoabsorption. Sedona handles compositional changes due to radioactive decay, decaying the initial distribution at mapping to FLASH to the
composition that should be present at that time.
The calculations assumed local thermodynamic equilibrium (LTE) to compute the ionization/excitation state of the gas.  The temperature in each zone of ejecta was calculated self-consistently by balancing the radiative heating and cooling. The opacity and emissivities were calculated based on the relevant radiative processes, including electron scattering, bound-free, free-free, and bound-bound transitions, the last of these treated in the expansion opacity formalism. 

\section{Results}\label{sec:results}

\begin{figure*}
    \centering
    \includegraphics[width=0.47\textwidth]{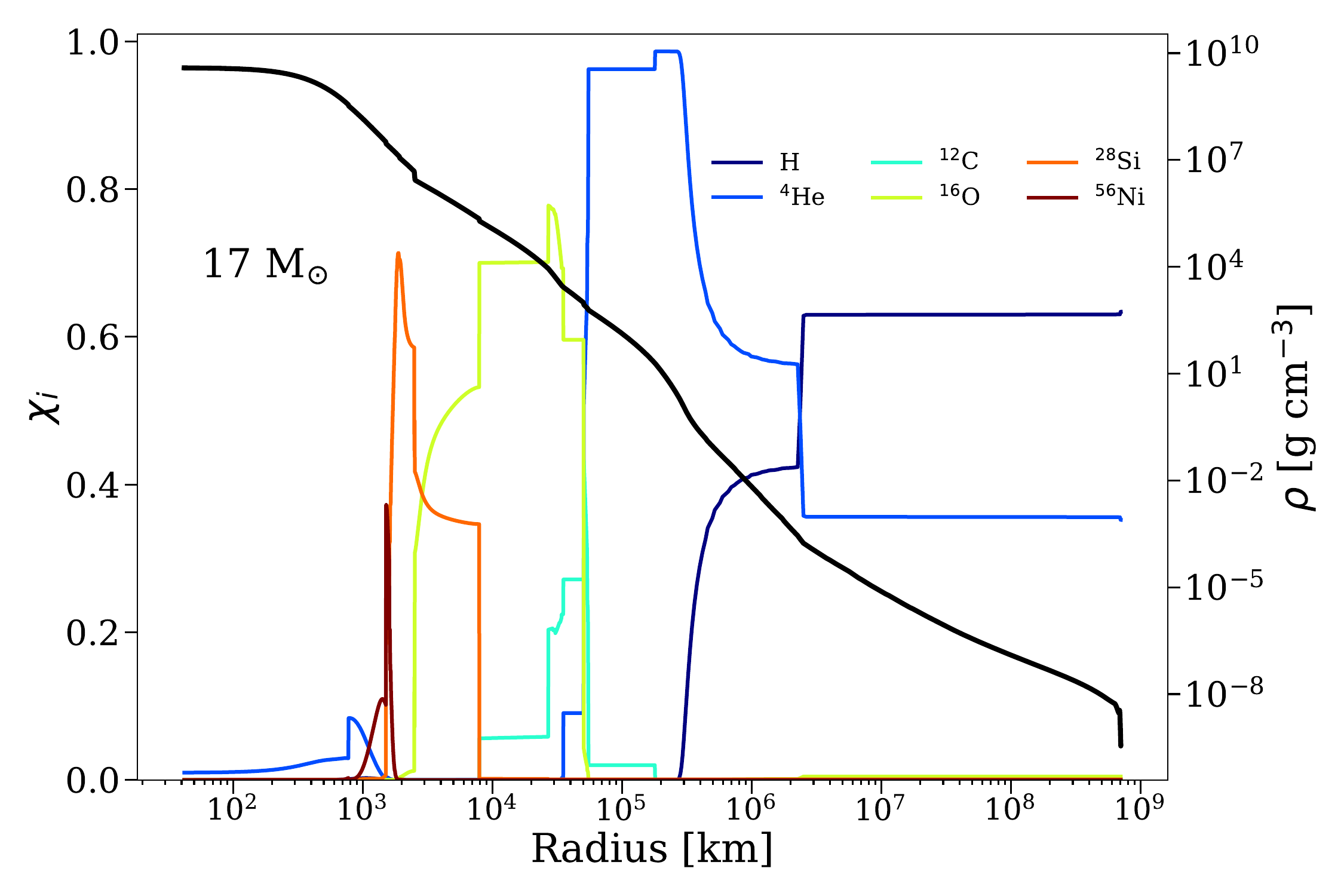}    
    \caption{Initial density profiles and composition for several key isotopes of the 17-M$_{\odot}$ progenitor model at the onset of core collapse, plotted against radial (km) coordinates. The left vertical axis shows the mass fraction of various elements, the right vertical axis the density (black line, g cm$^{-3}$). Compare with the first panel of Fig.\,\ref{fig:rho_iso} the composition and density profile at the time of mapping from F{\sc{ornax}} to FLASH.} 
    \label{fig:rho_init}
\end{figure*}

We illustrate the initial conditions at the onset of core collapse for the 17-M$_{\odot}$ model in Fig.\,\ref{fig:rho_init}, showing profiles of the density and the chemical compositions in radial coordinates. The C+O/He interface lies at $\sim$55,000\,km, and the H/He interface at $\sim$2.5 million km.

\begin{figure*}
    \centering
    \includegraphics[width=0.47\textwidth]{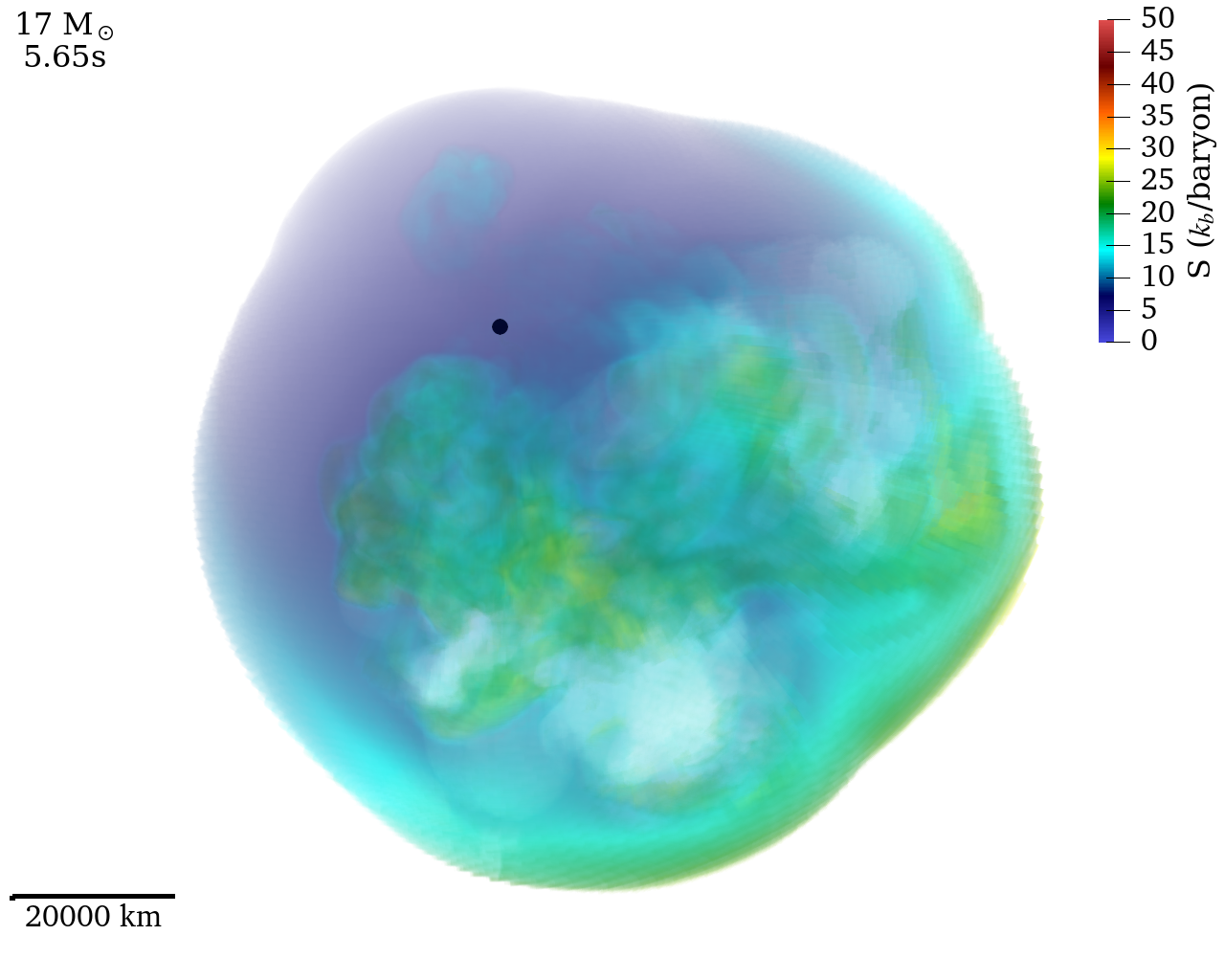}
    \includegraphics[width=0.47\textwidth]{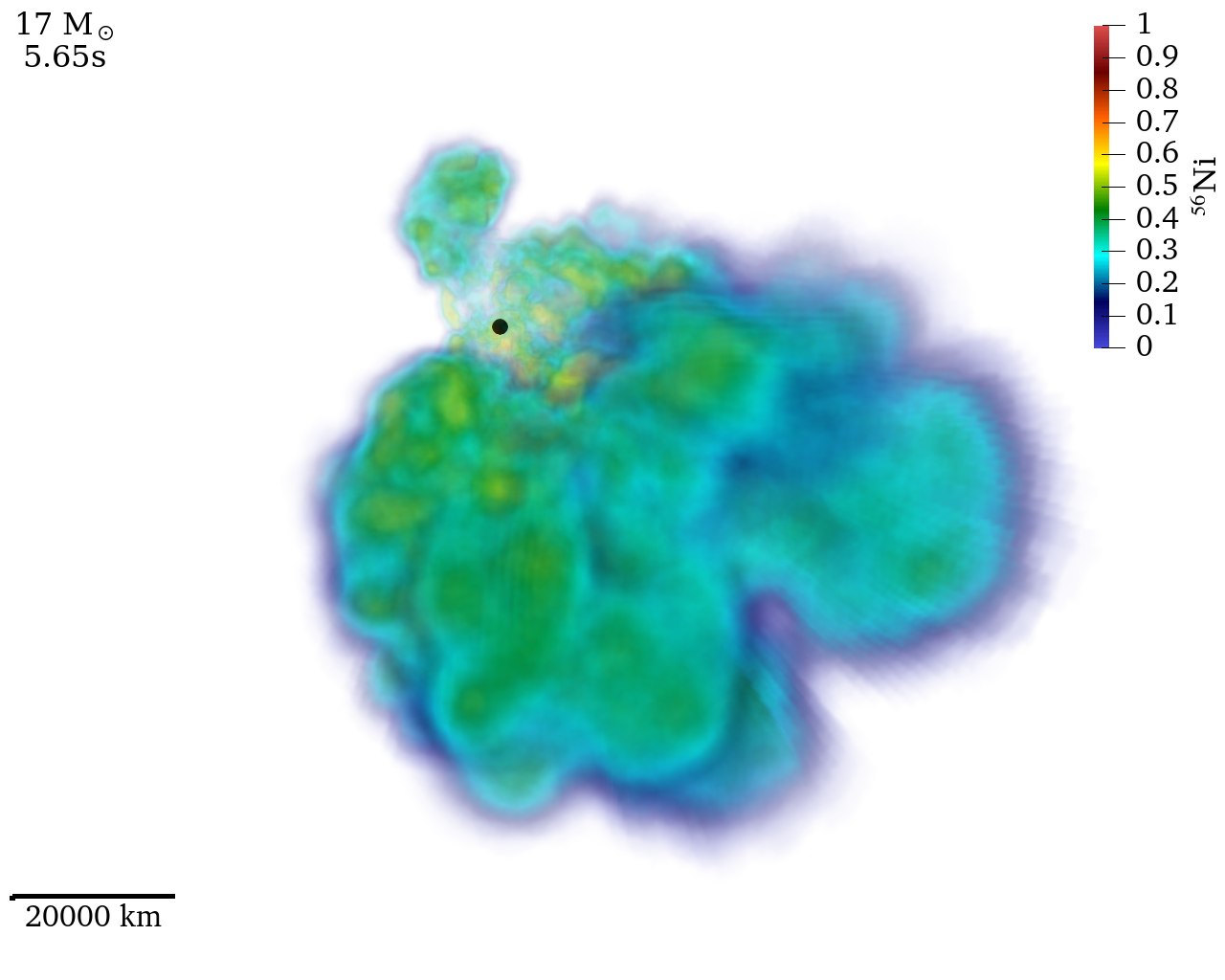}
    \caption{(\textbf{Left}) 3D volume renderings of entropy in the F{\sc{ornax}} simulation near the time of mapping into FLASH, at $\sim$5.56 s post-bounce, illustrating the shock surface. The PNS is represented as a black dot. We highlight the mild asymmetry in the shock surface as well as the highly asymmetric high-entropy plumes just interior. These plumes are where the dominant nickel formation occurs (\textbf{right}), with a large dipolar outflow crowned by a smaller `top-hat' of nickel. We emphasize here and throughout the similarity between the nickel structure at these early seconds and the structure at breakout, more than a day later, shown in Fig.\,\ref{fig:nivel}.}
    \label{fig:3D_init}
\end{figure*}

In Fig.\,\ref{fig:3D_init}, we show volume renderings of the specific entropy, indicative of the shock surface, and the nickel distribution at the end of the simulation in F{\sc{ornax}}, $\sim$5.56 seconds post-bounce, with the shock already past the C+O/He interface. The shock surface gently deviates from spherical symmetry and is perturbed by large-scale, highly asymmetric high-entropy plumes. Nickel formation occurs dominantly along these plumes, just interior to the shock. We state our conclusions first: the nickel distribution at the time of mapping bears strong resemblance to the distribution at shock breakout (Fig.\,\ref{fig:nivel}), with additional structure imparted by the reverse shock and nascent RTI. These results couple seconds to days, and kilometers to AUs, with shock and nickel structures that can be preserved out into the CSM and ISM \citep{gabler,orlando-cent,orlando-mil}.

\begin{figure*}
    \centering
    \includegraphics[width=0.47\textwidth]{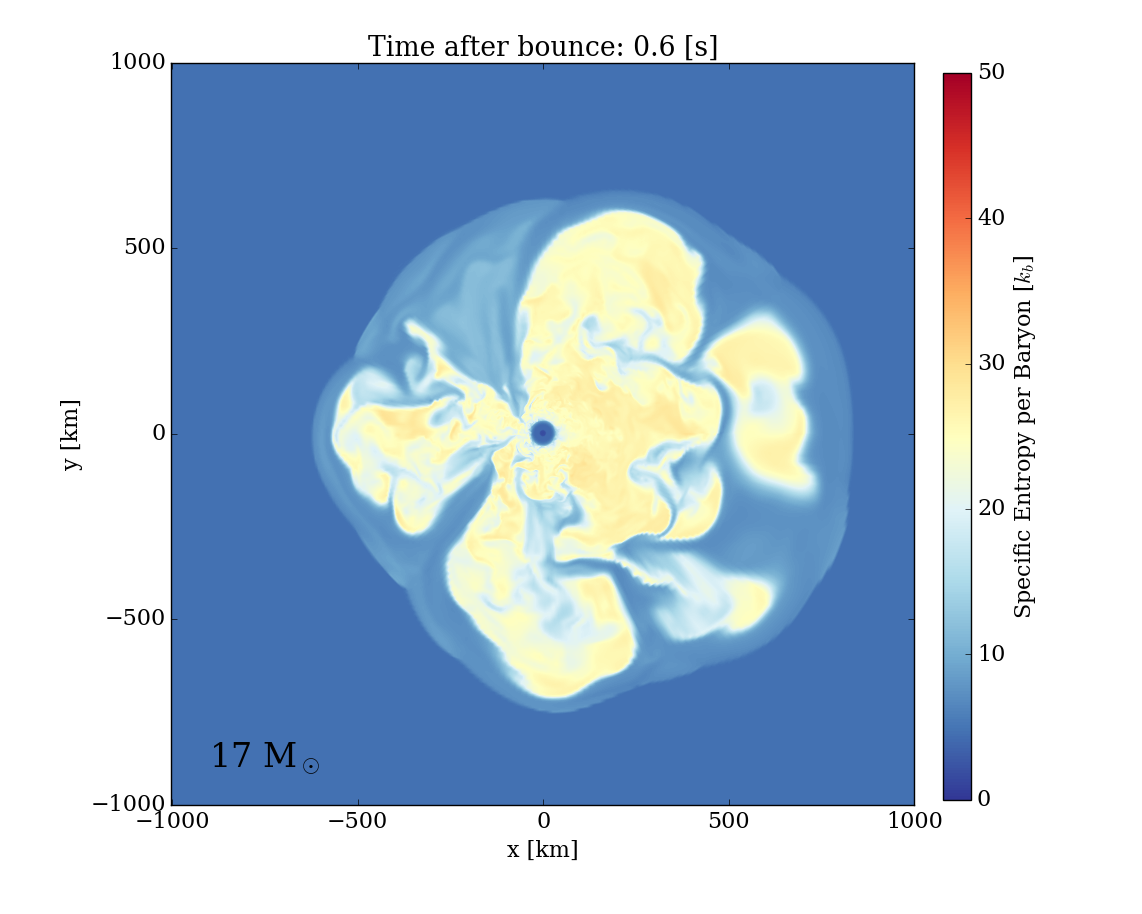}
    \includegraphics[width=0.47\textwidth]{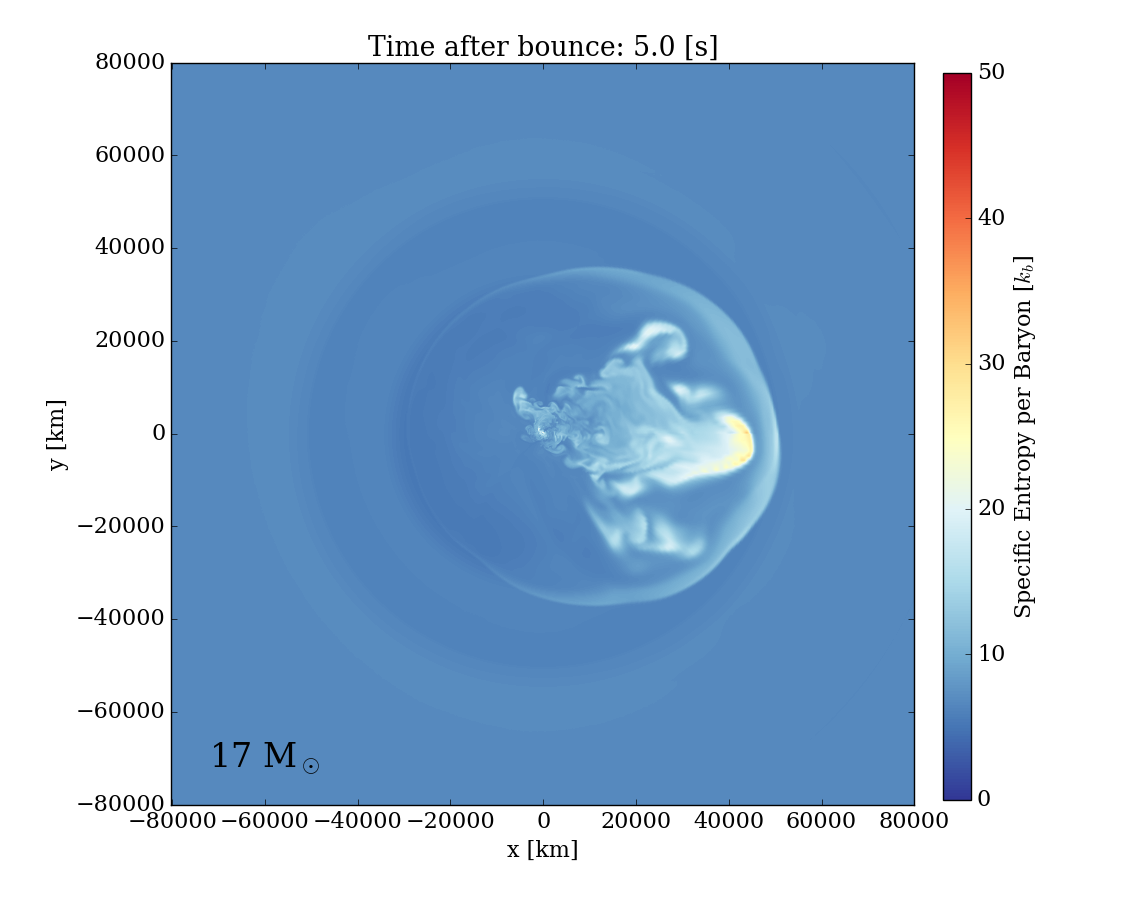}
    \includegraphics[width=0.47\textwidth]{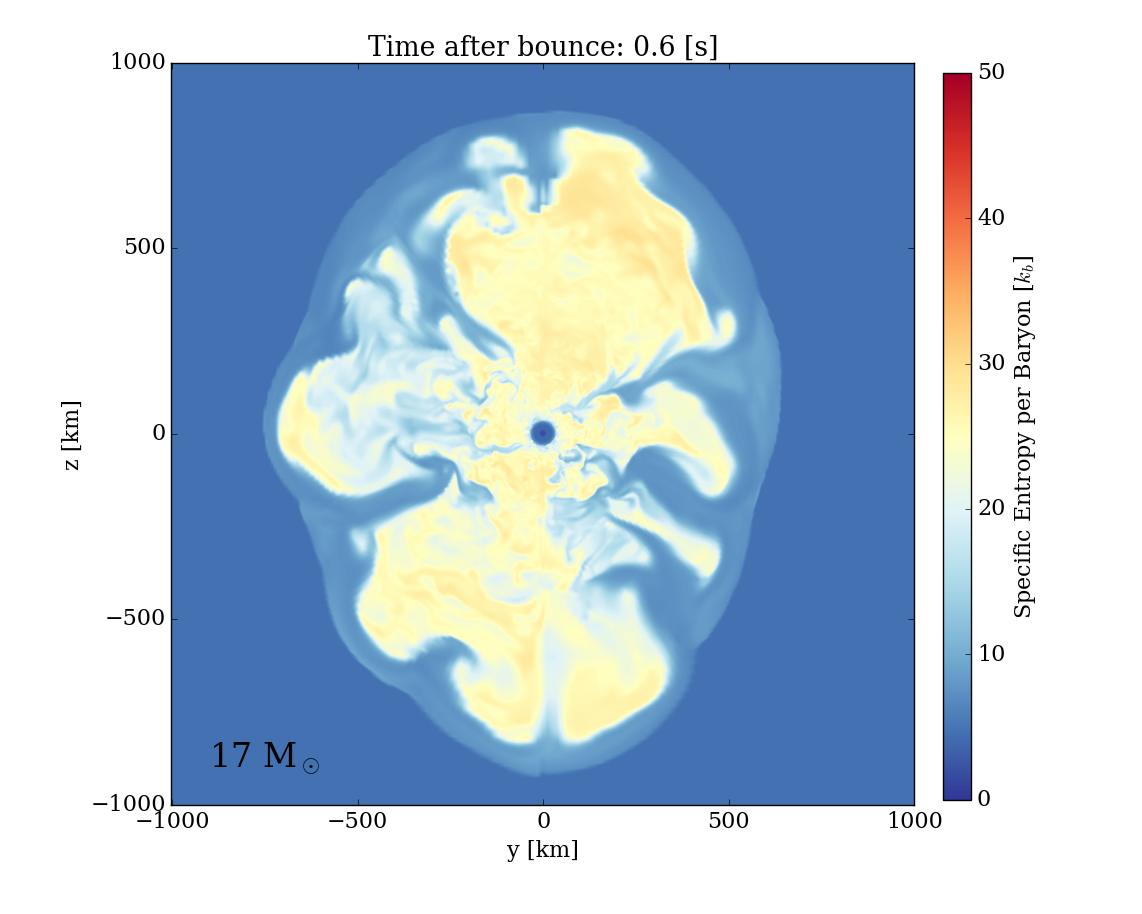}
    \includegraphics[width=0.47\textwidth]{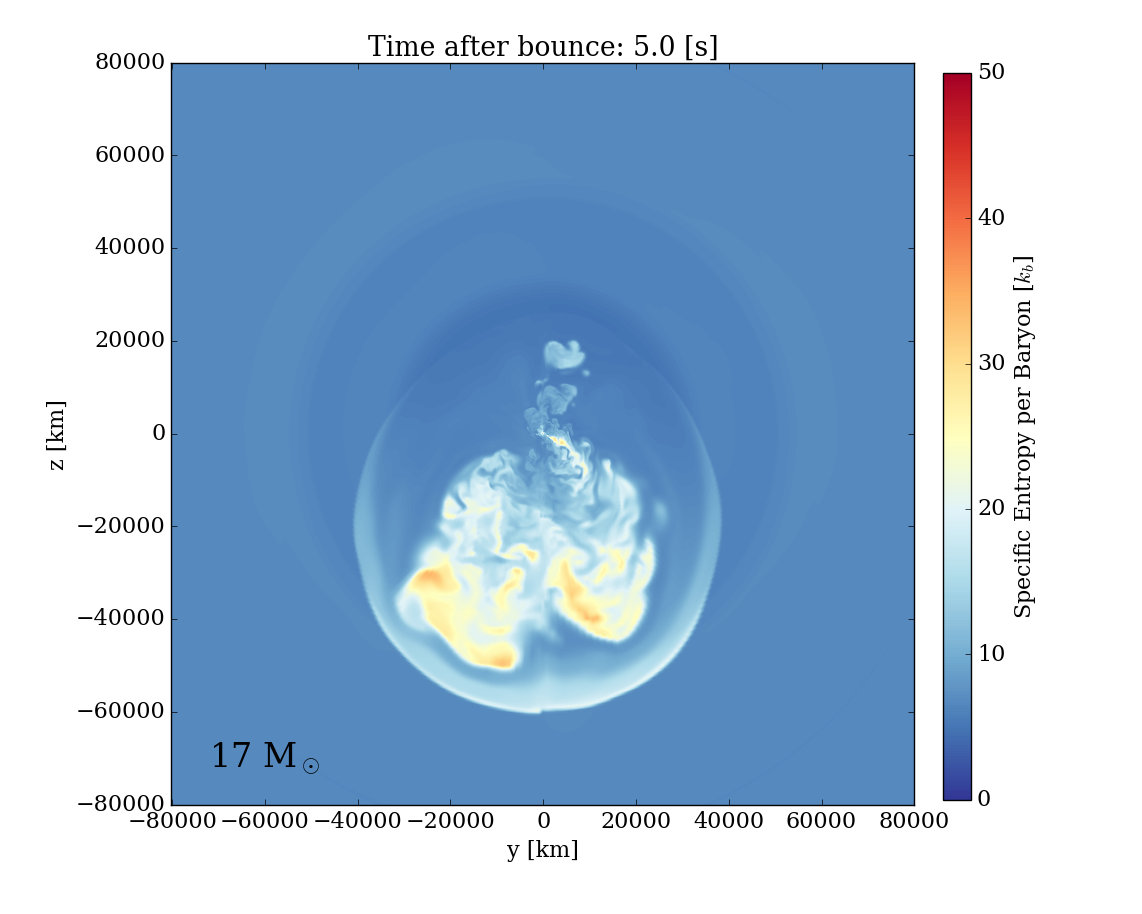}
    \includegraphics[width=0.47\textwidth]{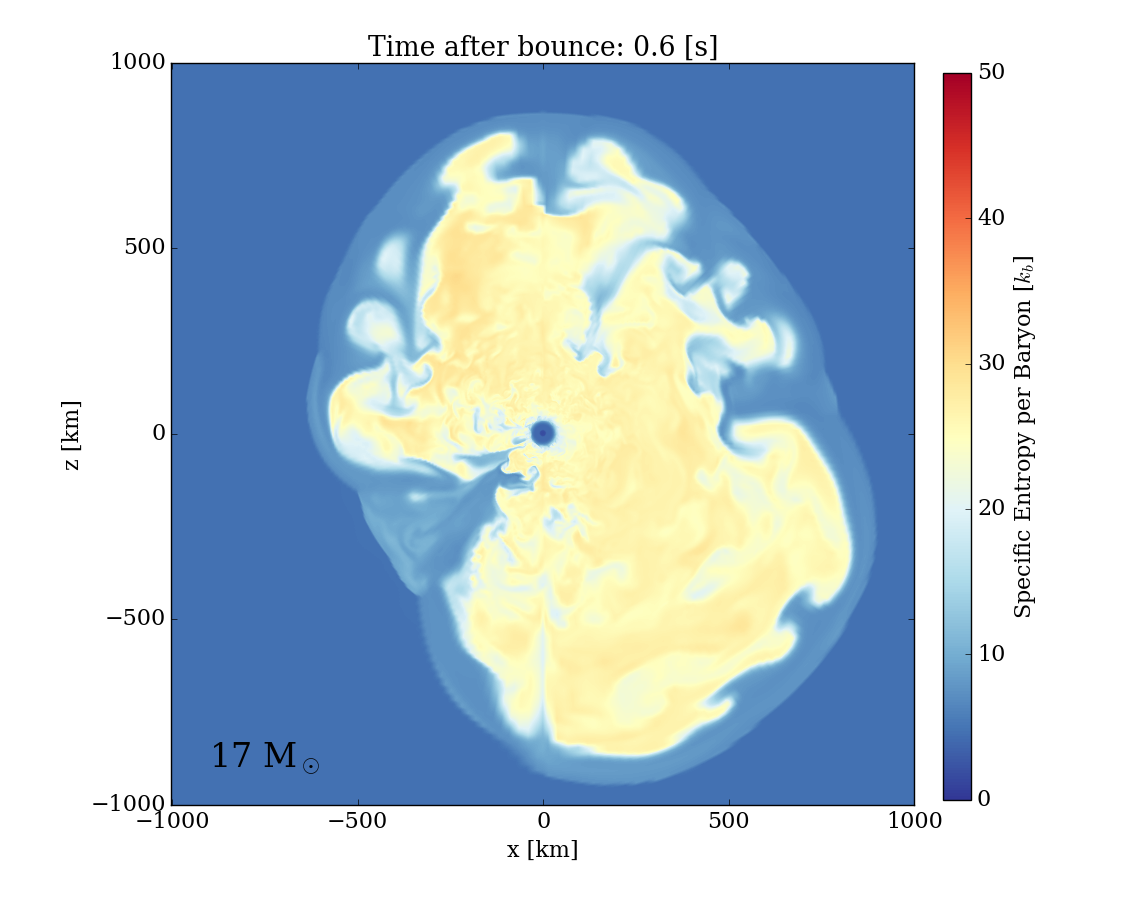}
    \includegraphics[width=0.47\textwidth]{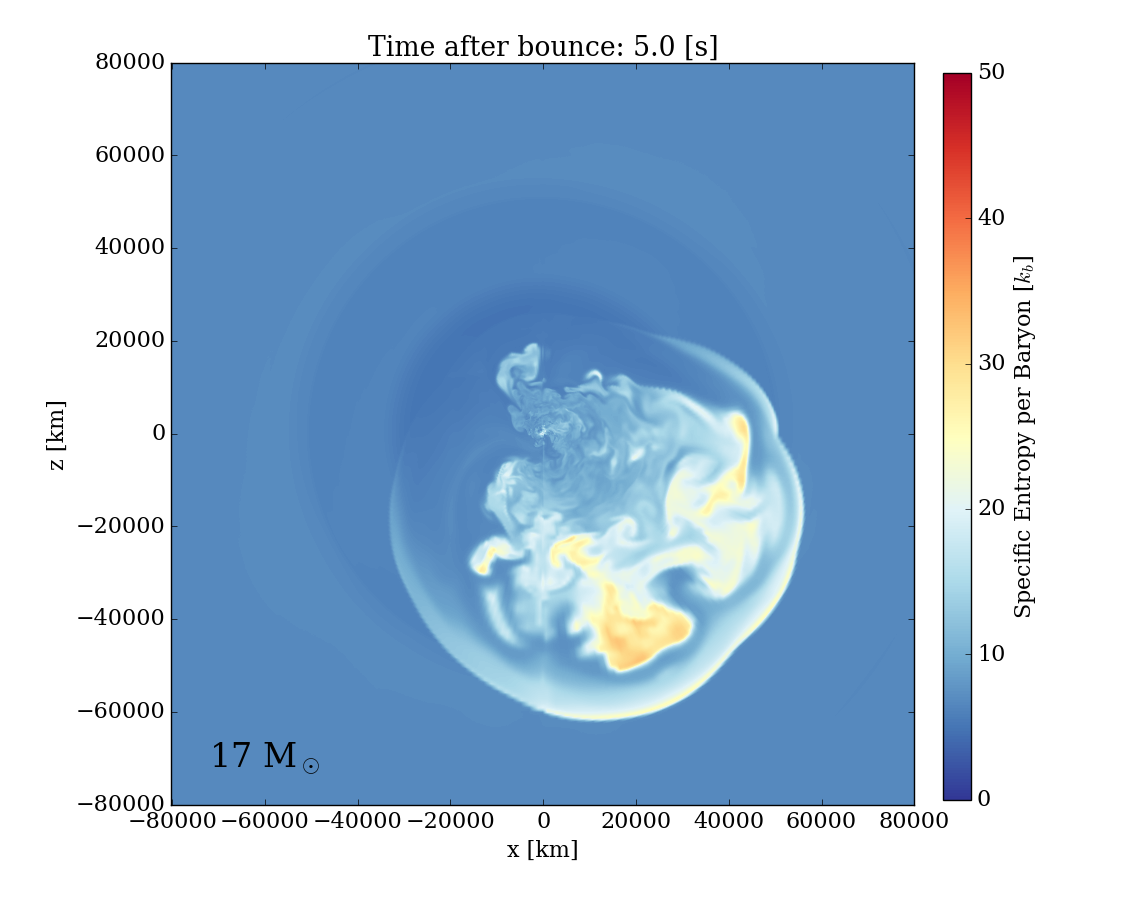}

    \caption{Entropy slices illustrating the shock evolution in F{\sc{ornax}} along three planes: an equatorial slice in x-z,  y-z, and x-z (\textbf{top} to \textbf{bottom}) at $\sim$0.6 s (\textbf{left}, just after shock revival and the launch of explosion) and at 5 s (\textbf{right}), before mapping into FLASH. Note the changing scales. Compare the stark similarity in the entropy structure in the bottom right panel at $\sim$5 s with the nickel distribution in Fig.\,\ref{fig:nivel}, at $\sim$119,000 s after core bounce. }
    \label{fig:dens_init}
\end{figure*}

Fig.\,\ref{fig:dens_init} illustrates the entropy evolution in F{\sc{ornax}} just after shock revival (marking the onset of explosion), at $\sim$0.6 seconds, and at $\sim$5 seconds after core bounce, just before mapping to FLASH. The initial multipolar structure, consisting of several large plumes, abates as plumes accrete and coalesce. The plume structure will continue to evolve through fallback accretion, but the final shock and nickel distributions at breakout mimic strongly those at mapping. For our energetic, asymmetric 17-M$_{\odot}$ model, the structural asymmetries shaped by neutrino-driven turbulence on second-timescales are frozen out and preserved through the end of our simulation, beyond five days.

\subsection{Terminology}

We briefly define the jargon used in existing literature to describe the explosion geometry. We categorize \textit{plumes} as large-scale structures formed by neutrino-driven heating and expanding into the stellar envelope. Within the plume, we see the formation of \textit{RTI fingers}, smaller-scale extended distributions containing nickel and other metals that form upon interaction with the reverse shock at the H/He interface, defined at the radial and mass coordinate where the angle-averaged mass fractions of $^4$He and H are equal. By \textit{bullets}, we specifically mean parcels of nickel/other metals, characteristically dense and energetic, moving at high-velocities into this interface. Lastly, we refer to \textit{clumps} (often used interchangeably with fingers/bullets, \citealt{gabler}) to indicate again smaller-scale parcels of metal, usually nickel, that show spatial correlation without reference to their velocities.

\subsection{Reverse Shock Formation and RTI}

\begin{figure*}
    \centering
    \includegraphics[width=0.44\textwidth]{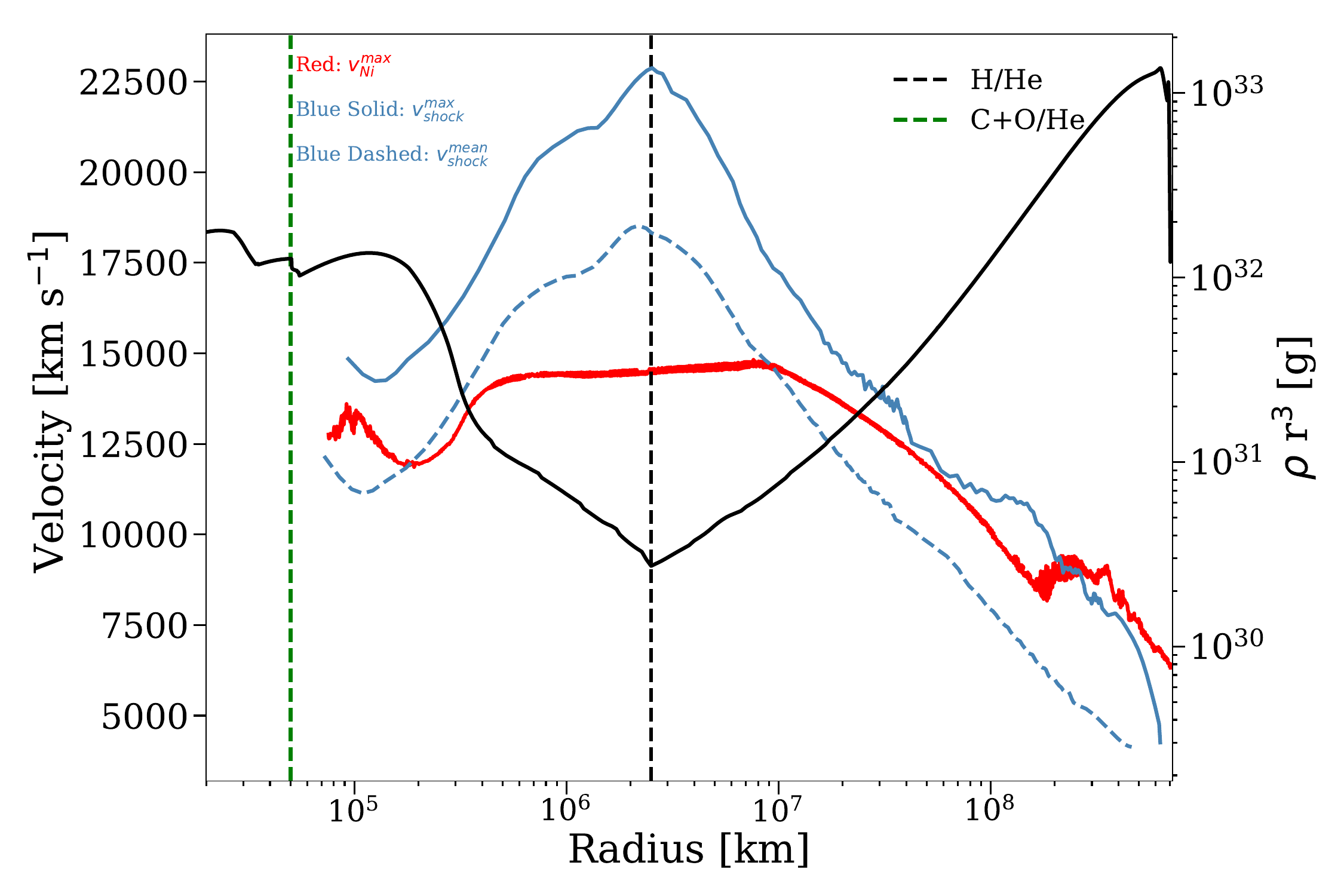}
    \includegraphics[width=0.47\textwidth]{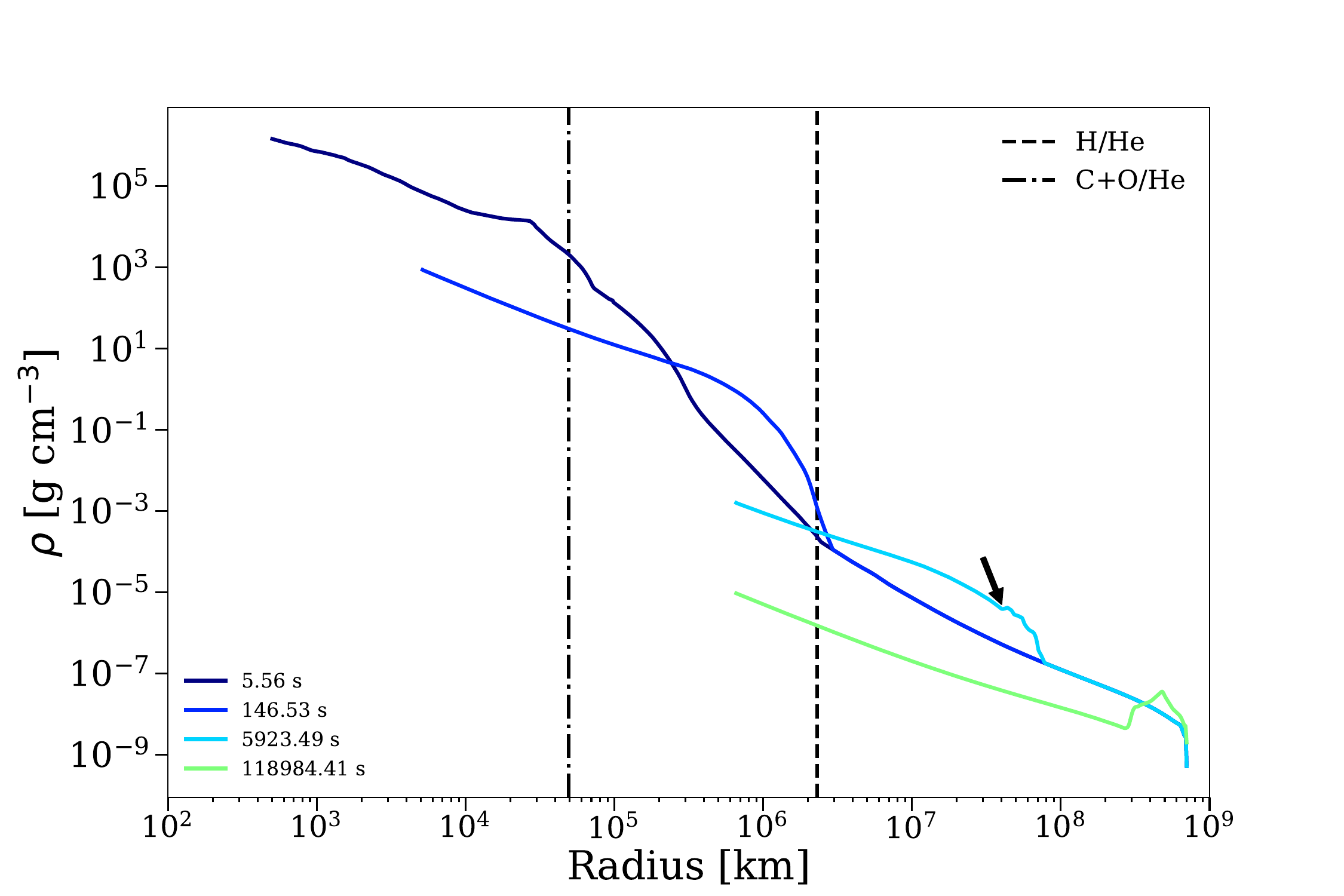}
    \caption{ \textbf{Left:} The evolution of the mean and maximum velocity of the shock surface (blue), the maximum nickel velocity (red, in km s$^{-1}$, left vertical axis), and the core collapse density profile $\rho\,r^3$ (g cm$^{-3}$, black, right vertical axis) all plotted against radius (km). The shock reaches peak velocities of $\sim$23,000\,km\,s$^{-1}$ going down the density gradient until it arrives at the H/He interface at $\sim$146 s, at $\sim$2.5 million km. The shock then decelerates until breakout to several 1000\,km\,s$^{-1}$. Not shown is the subsequent uptick of the shock velocity in the photosphere, where the the density plummets. \textbf{Right:} Angle-averaged density (in g\,cm$^{-3}$) profile plotted against radius (in km) for various time snapshots. The vertical lines show the H/He and C+O/He interfaces. At $\sim$146\,s, the shock first encounters the H/He interface. A reverse shock is visible in the bifurcation of the density profile by $\sim$6000\,s, indicated by the arrow, but continues to propagate outward in radius for several days.}
    \label{fig:dens_avg}
\end{figure*}

A shock expanding adiabatically, absent radiative loss, conserves energy during its hydrodynamic evolution. As the shock expands and sweeps up material, its velocity will decrease in response to the mass pile up and there will be an interplay between the kinetic and internal components of the explosion energy, ultimately trigger a reverse shock. The left panel of Fig.\,\ref{fig:dens_avg} shows this trend, plotting the radial evolution of the shock and nickel velocities (red and blue, respectively)  together with the radial profile of the angle-averaged density (black), $\rho$\,r$^3$, at core bounce.  The shock begins exterior to the C+O/He core and experiences a brief initial deceleration after mapping due to the uptick in  $\rho$\,r$^3$. 

While within the helium core, the shock accelerates down the density gradient, reaching peak velocities of $\sim$23,000 km\,s$^{-1}$, before it reaches the H/He interface. At the H/He interface, a reverse shock is formed which propagates inward in mass coordinates towards the metal-rich ejecta. The forward shock continues to move outward and decelerates as it sweeps up the mass of the hydrogen envelope. We identify hydrodynamic shock breakout as the first emergence of the asymmetric shock from the stellar surface, which occurs at a time $\sim$92,000 seconds post-bounce with a shock velocity  $\sim$5000 km\,s$^{-1}$.

The right panel of Fig.\,\ref{fig:dens_avg} illustrates this qualitative behavior with snapshots of the angle-averaged density profile for four times: at mapping from F{\sc{ornax}} to FLASH $\sim$5.56\,s post-bounce; at the first shock intrusion into the H/He interface $\sim$146 seconds; at interaction with the reverse shock thereafter ($\sim$6000\,s); and at shock breakout along the equatorial direction($\sim$119,000\,s). 
The shock is visibly identifiable as the sharp density discontinuity. The inner boundaries of the curves shift to higher radii as we excise more of the innermost cells on the simulation grid. The density over the modeled domain drops by over 15 orders of magnitude through the course of the simulation. The shock crosses the H/He interface at $\sim$146 seconds, but a reverse shock only forms at $\sim$6000 seconds, indicated by an arrow at the bifurcation in the density profile at the forward shock,
as sufficient matter is swept up and the kinetic energy of the explosion is funneled into internal energy. The reverse shock continues moving outward in radius, but inward in mass, through shock breakout, although the radial separation between forward and reverse shock increases. We followed our simulation until $\sim$ one week after core bounce, when the reverse shock just reaches the stellar interior. 

\begin{figure*}
    \centering
    \includegraphics[width=0.47\textwidth]{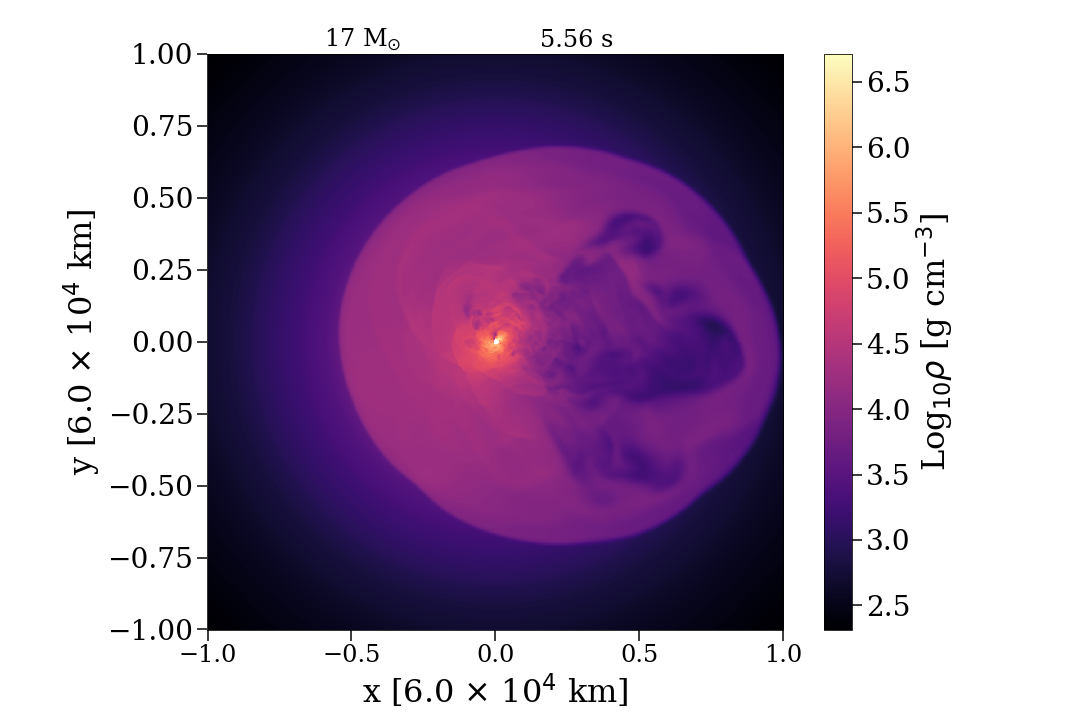}
    \includegraphics[width=0.47\textwidth]{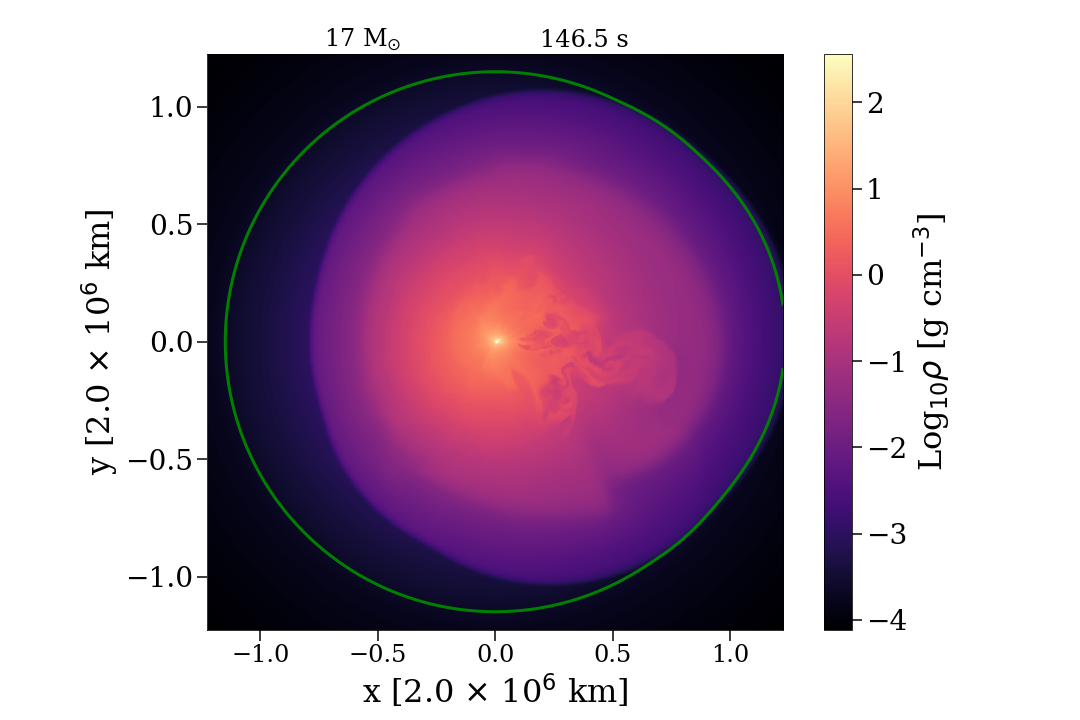}
    \includegraphics[width=0.47\textwidth]{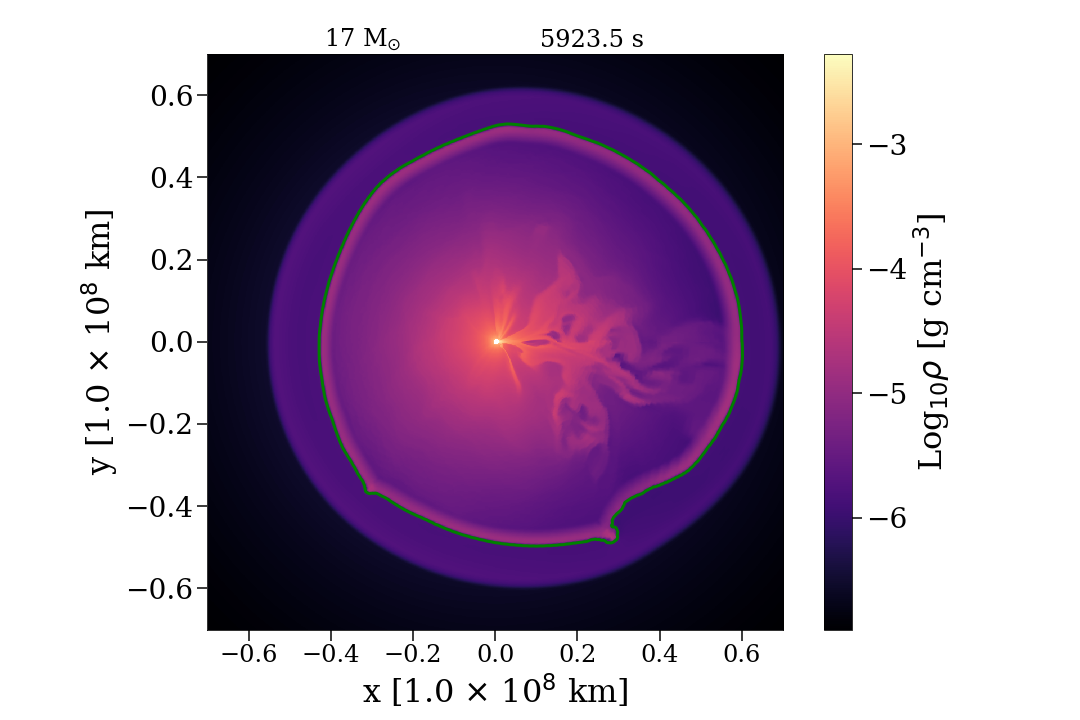}
    \includegraphics[width=0.47\textwidth]{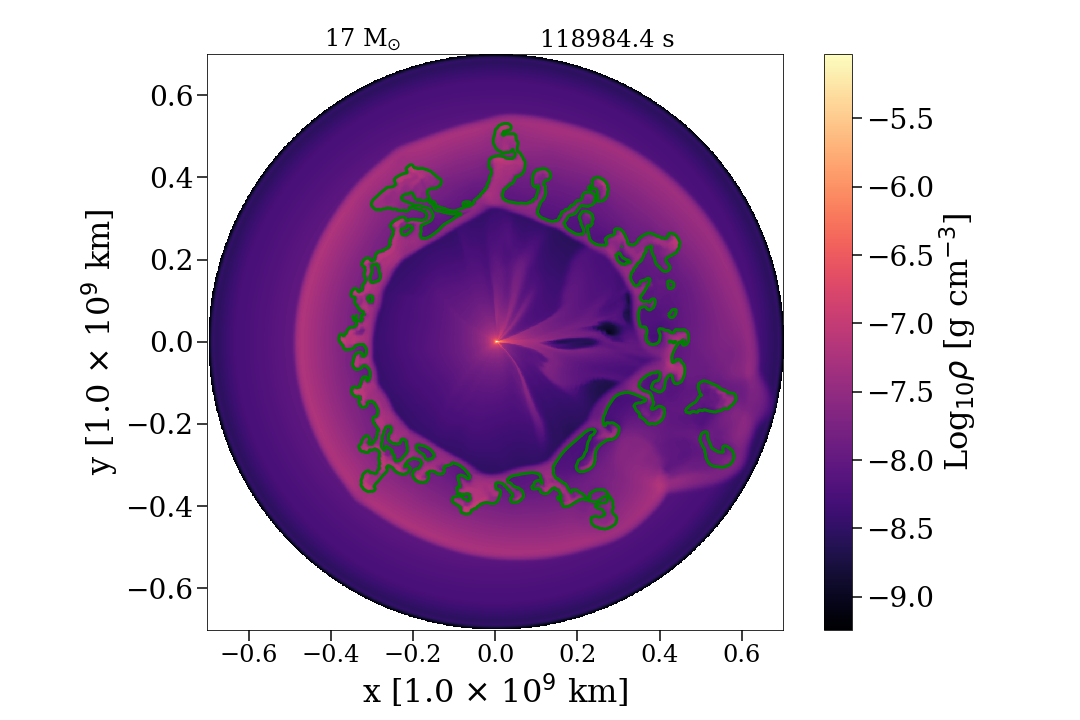}
    
    \caption{Density slices (in log$_{10}$ g\,cm$^{-3}$) along the equatorial plane at various times (at mapping, 5.65 s post-bounce, when the shock arrives at the H/He interface at $\sim$146 s, the formation of a reverse shock $\sim$6000 s, and at $\sim$119,000 s, after shock breakout) showing the evolution of the shock. Both the grid and color bar scales vary with time. The green contour illustrates the H/He interface, defined where their mass fractions are equal. RTI-driven protrusions become visible through the interface by ${\sim}10^{4}$\,s.}
    \label{fig:dens}
\end{figure*}

We highlight snapshots of the density profile along the equatorial plane in Fig.\,\ref{fig:dens} for a similar time sequence as in the right panel of Fig.\,\ref{fig:dens_avg}. The shock front, visible as the sharp density contrast, is elliptical in shape. The plumes just interior to the shock identify the nickel entrained by the shock. After the shock hits the H/He interface (green contour on the top right panel), we see the separation grow between the forward moving shock and the reverse shock demarcated by this interface.  Coincident with reverse shock development, the RTI manifests as nickel ejecta carve out excursions along multiple directions through the hydrogen-helium interface and into the hydrogen envelope, discussed in more detail below. 
At breakout, the reverse shock trails the outward shock by $\sim$3$\times$10$^{8}$ km, with some RTI fingers spanning nearly the entire width, as seen in the bottom right panel of Fig.\,\ref{fig:dens}. We emphasize the high degree of asymmetry in the explosion. Shock breakout occurs along the southern hemisphere at $\sim$92,000\,s. Breakout along the northern hemisphere occurs almost one day later, at $\sim$170,000\,s, and at $\sim$118,000\,s along the equatorial direction. Such an aspherical shock breakout would have direct observational consequences, smearing out the initial breakout flash \citep{2021MNRAS.508.5766I}. This result would appear degenerate with a denser CSM distribution. The direction of first shock breakout aligns with the dominant axis of asymmetry in the early-time shock structure within the F{\sc{ornax}} simulation. This is visible in the equatorial plane by comparing the bottom right and the top left panels, with the most extended shocked plume preserving its direction through breakout, coupling vastly different time and length scales.

\subsection{Nickel Evolution}

\begin{figure*}
    \centering
    \includegraphics[width=0.9\textwidth]{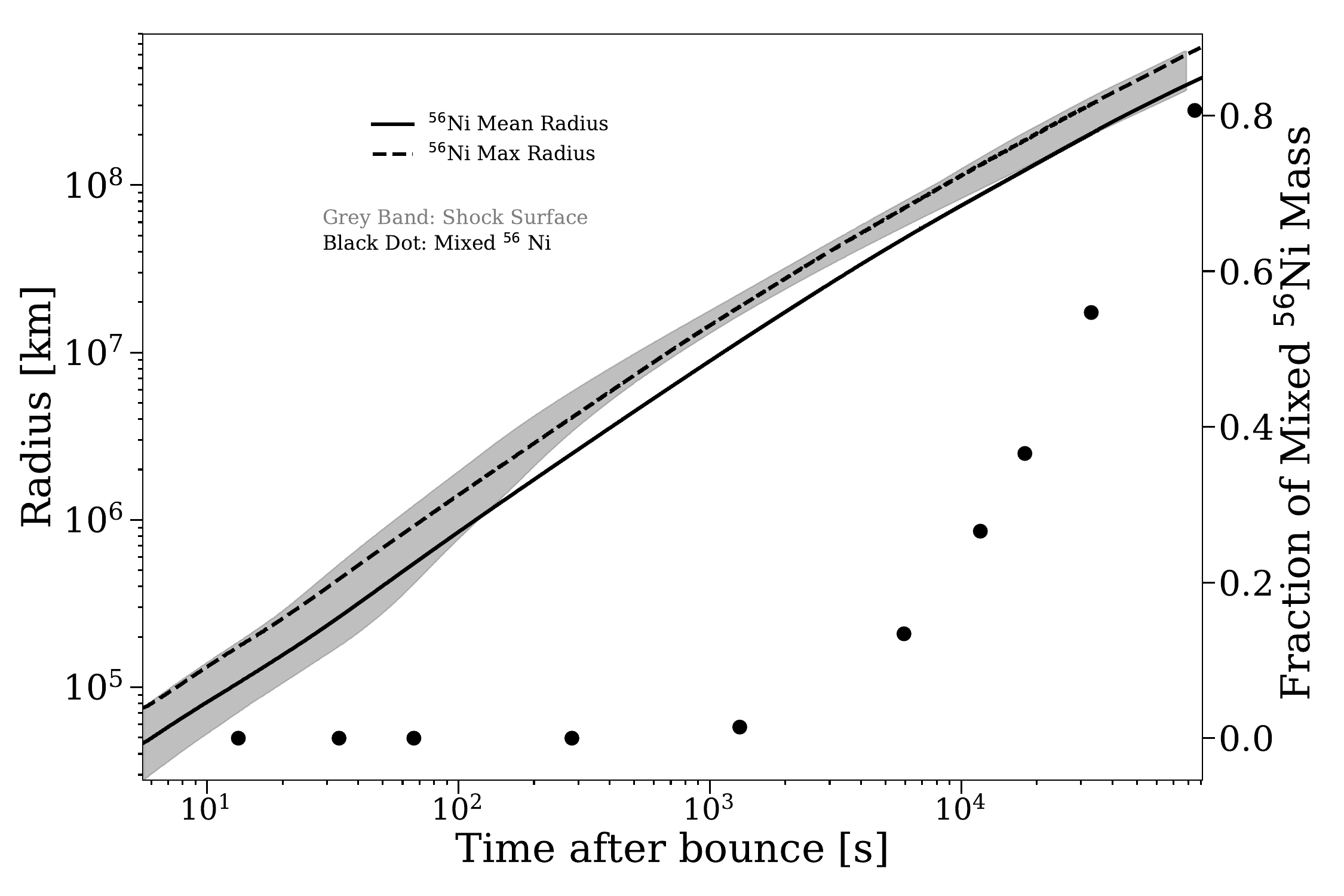}
    \caption{Time evolution of the radial extent of the shock surface (gray band) and the $^{56}$Ni ejecta trajectory (black lines). We show also the time evolution of the fraction of nickel mass mixed outward into the hydrogen envelope, beyond the H/He interface, as black dots. At mapping, the nickel, which is formed just interior to the shock surface, is just cresting the shock. As the shock goes down the density gradient and accelerates, the nickel bullets trail further behind the shock. After the shock hits the H/He interface at $\sim$146 and decelerates, the nickel continues to coast inertially for $\sim$6000 s, catching up with the shock. Simultaneously, we see a marked rise in the nickel mixing $-$ by shock breakout, the majority of the nickel has mixed outwards, though it has not penetrated through the forward moving shock surface. Beyond this point, nickel no longer will puncture through the shock as the shock now accelerates down the density gradient into the CSM.}
    \label{fig:bullet}
\end{figure*}

The nickel evolution from bounce through breakout follows the shock trajectory, with key differences. Our 17-M$_{\odot}$ model had $\sim$0.1\,M$_{\odot}$ of $^{56}$Ni at breakout. We show the nickel velocity vs time in the left panel of Fig.\,\ref{fig:dens_avg} and its trajectory vs time in Fig.\,\ref{fig:bullet}. At the time of mapping, the nickel lags just behind the outermost shock front, ahead of the angle-averaged shock radius but interior to the maximum shock surface. The initial wiggles in its velocity reflect the nature of the density structure at the C+O/He interface. While the shock accelerates through the helium core, the nickel generally coasts at a constant velocity and falls further behind the shock front for the first hour. Though the shock first encounters the H/He interface at $\sim$2.5 million km, the fastest nickel parcels only begin to decelerate once they have expanded eight-fold further to $\sim$20 million km, reaching the H/He interface now swept outward by the shock.  By this point, the nickel has surpassed the mean shock radius and closely tails the maximum shock surface. Another tenfold further out in radius, at $\sim$200 million km, significant mixing occurs and the fastest-moving nickel moves along buoyant RTI-driven fingers to supersede the maximum shock surface velocity. However, the nickel does not penetrate the maximum shock front by breakout.  After breakout, the shock continues to accelerate down the lower density CSM, while the nickel structures lag further behind.  The maximum nickel velocity, just before deceleration by the reverse shock, is $\sim$14,500 km\,s$^{-1}$. For a stripped envelope progenitor $-$ which experiences earlier breakout  without an extended hydrogen mantle, a strong H/He interface, or a significant reverse shock $-$ we would expect similar velocities at breakout.

\begin{figure*}
    \centering
    \includegraphics[width=0.47\textwidth]{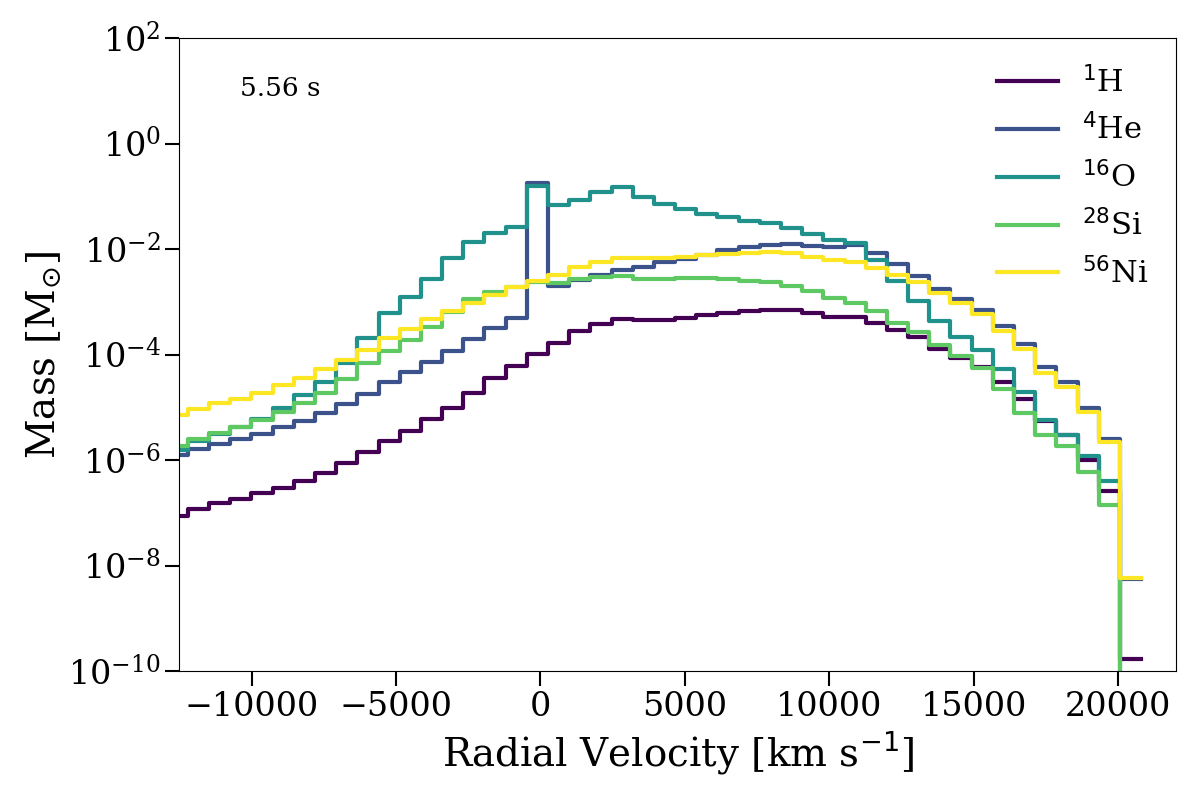}
    \includegraphics[width=0.47\textwidth]{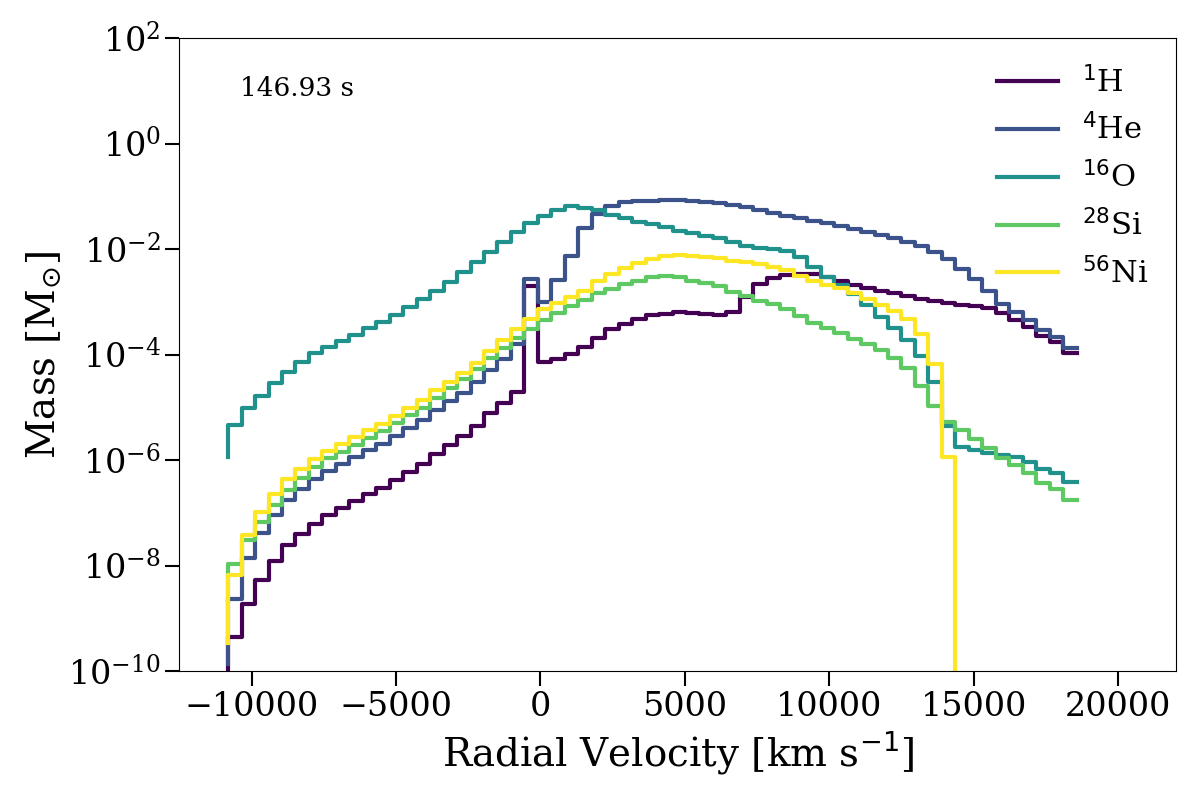}
    \includegraphics[width=0.47\textwidth]{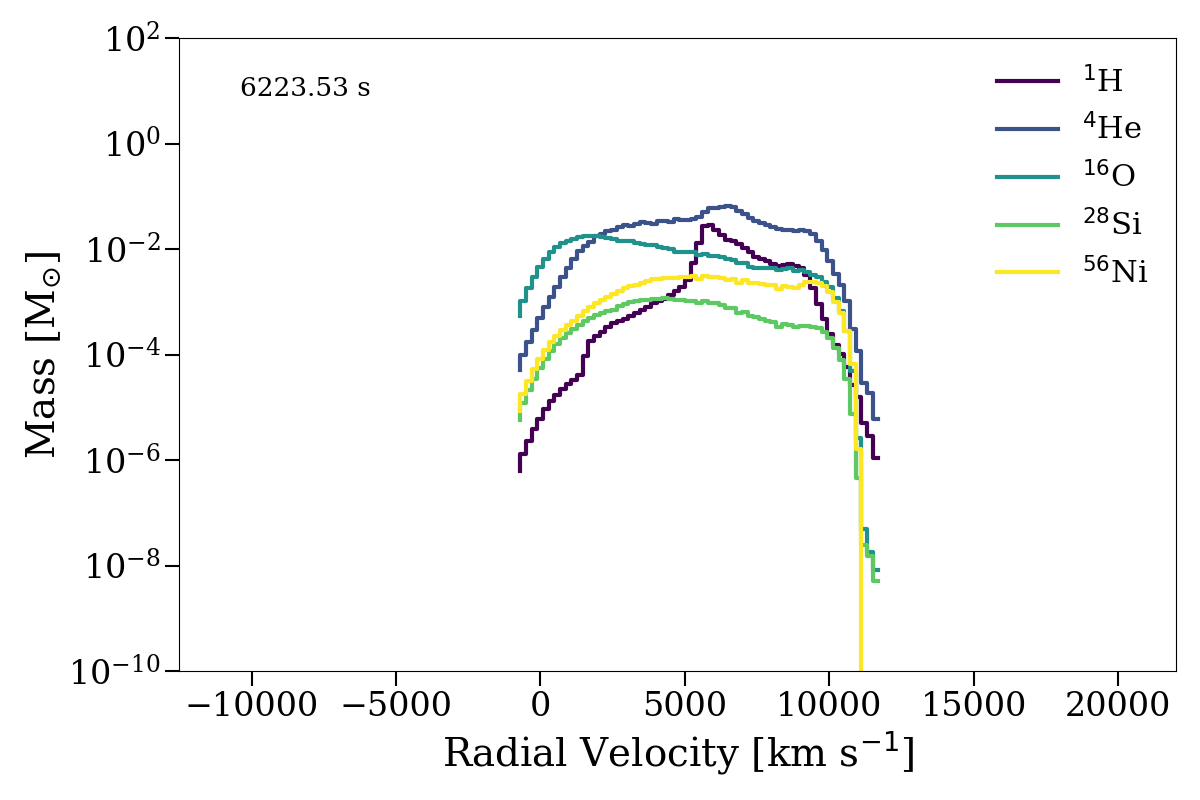}
    \includegraphics[width=0.47\textwidth]{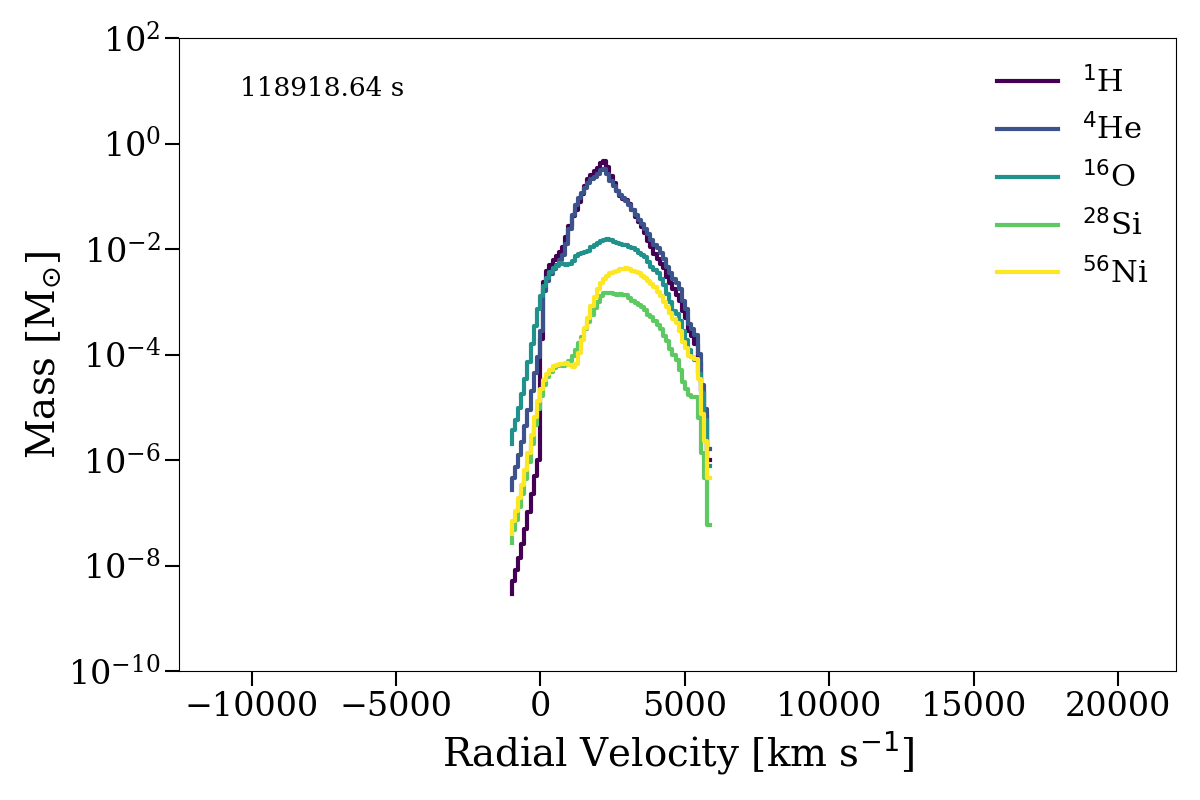}
    \caption{A histogram of isotope masses (M$_{\odot}$) by velocities (km s$^{-1}$) at various times for matter interior to the shock. At shock breakout, the bulk of $^{56}$Ni is moving at an average radial velocity of $\sim$3400\,km\,s$^{-1}$. The peak near 0\,km\,s$^{-1}$ for light metals, hydrogen, and helium is the result of the nearly-static KEPLER stellar envelope at mapping reflecting the asymmetry of the shock and vanishing as the shock overtakes it. }
    \label{fig:iso_hist} 
\end{figure*}

To further illustrate the composition velocity evolution,  we show histograms for various isotope mass fractions as a function of their radial velocity (km s$^{-1}$) for the familiar sequence of times in Fig.\,\ref{fig:iso_hist}. Between $\sim$140$-$10,000 seconds, the lighter elements exterior to nickel, swept along with the shock, accelerate and decelerate along the density gradients. Subsequently, the fastest nickel parcels, in the high-velocity tail, closest to the shock encounter the H/He interface and decelerate. The bulk of the nickel is unaffected, and its average velocity unperturbed, decelerating only as it rams into the H/He interface at $\sim$6000\,s. Coincidentally, the fastest moving nickel accelerates again, visible as the small hump in the histogram, with the formation of RTI.  $\sim$0.008 M$_{\odot}$ of the nickel ejecta at $\sim$146 s moves faster than 10,000 km\,s$^{-1}$.  At breakout, $\sim$0.001 M$_{\odot}$ of nickel is moving at velocities greater than 4800 km\,s$^{-1}$. Except for the small amounts of nickel at the highest mass fractions, generally, higher mass fraction parcels have higher velocities (and are located further out).

\begin{figure*}
    \centering
        \includegraphics[width=0.75\textwidth]{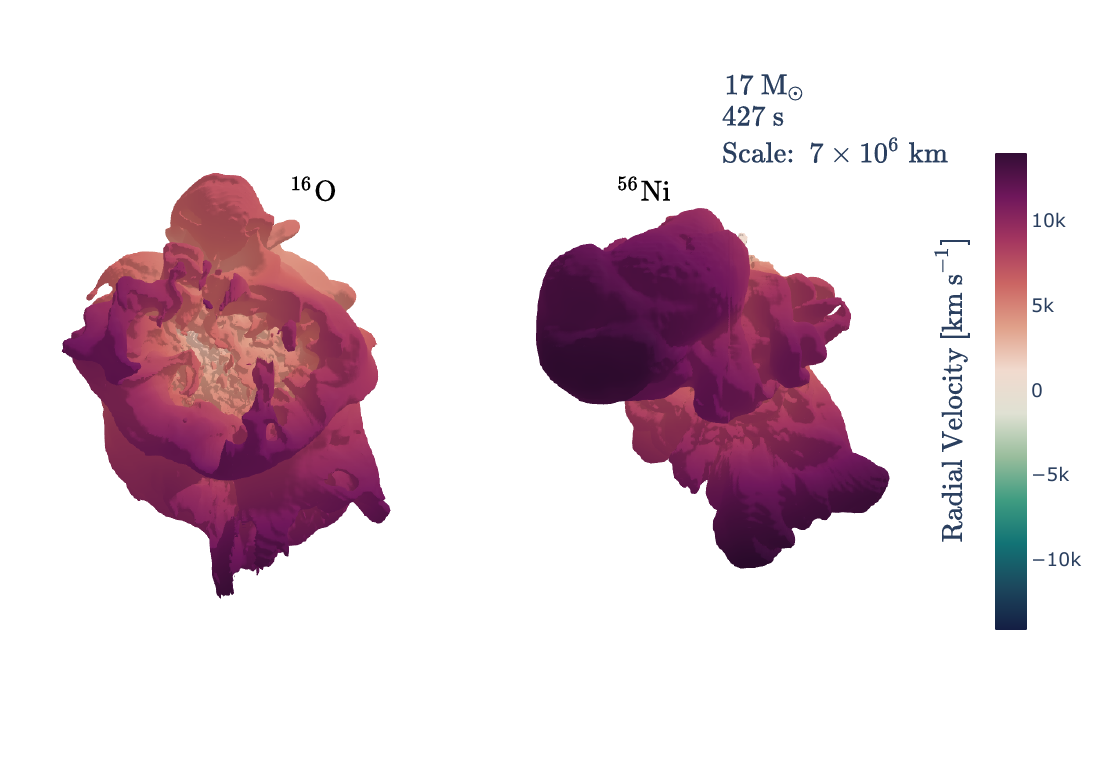}    \includegraphics[width=0.51\textwidth]{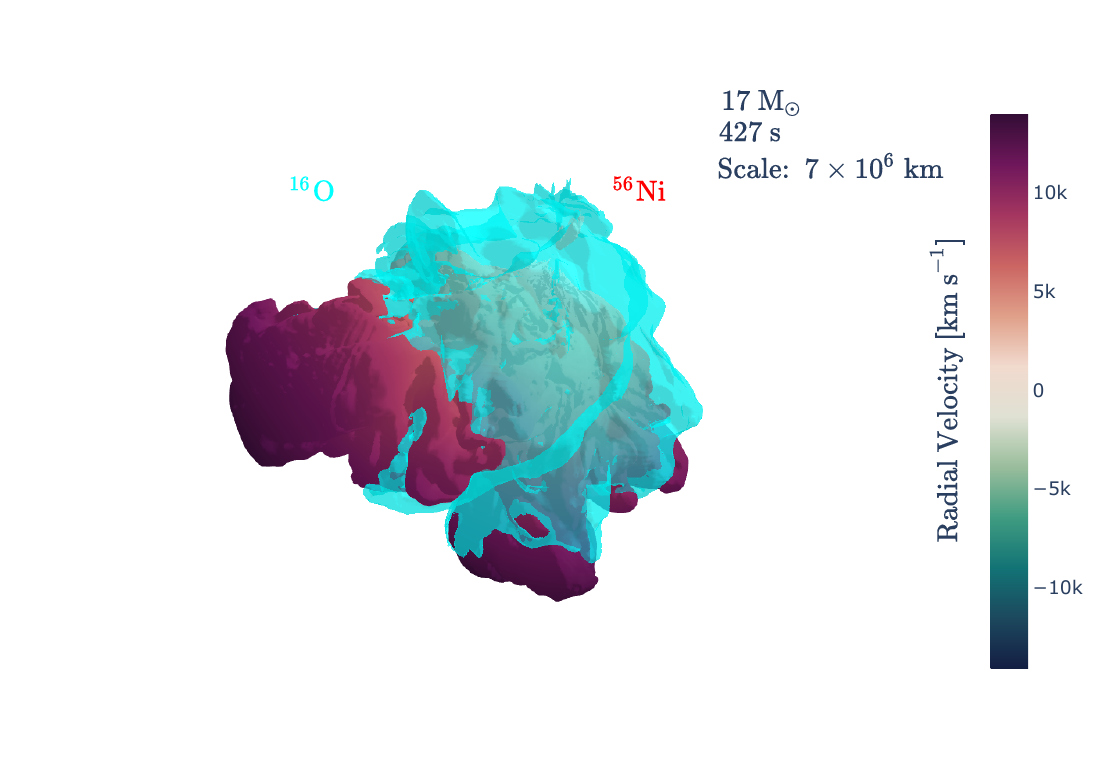}
    \caption{\textbf{Top} A 3$\%$ isosurface plot of oxygen (left) and $^{56}$Ni (right) at 427 s with a wingspan of 7 million km. In the nickel distribution, note the geometry with large and intermediate-scale plumes, and even a ring-like structure protruding in the  top right of the morphology. Smaller scale fingers and bullets emerge hours later, upon interaction with the reverse shock and the development of RTI. \textbf{Bottom}: The two isotopes overplotted, indicated by a blue veil and red surface, respectively. The nickel is colored by radial velocity, with the dominant plumes moving at $\sim$12,000 km s$^{-1}$. The nickel is formed in interior of the metal core and carves its way through the oxygen core as it breaks through. Thus, the structures of the two are complementary $-$ in a 2D slice, we would see the oxygen distribution transverse to the that of the nickel.}
    \label{fig:ni_over}
\end{figure*}

\begin{figure*}
    \centering
    \includegraphics[width=0.44\textwidth]{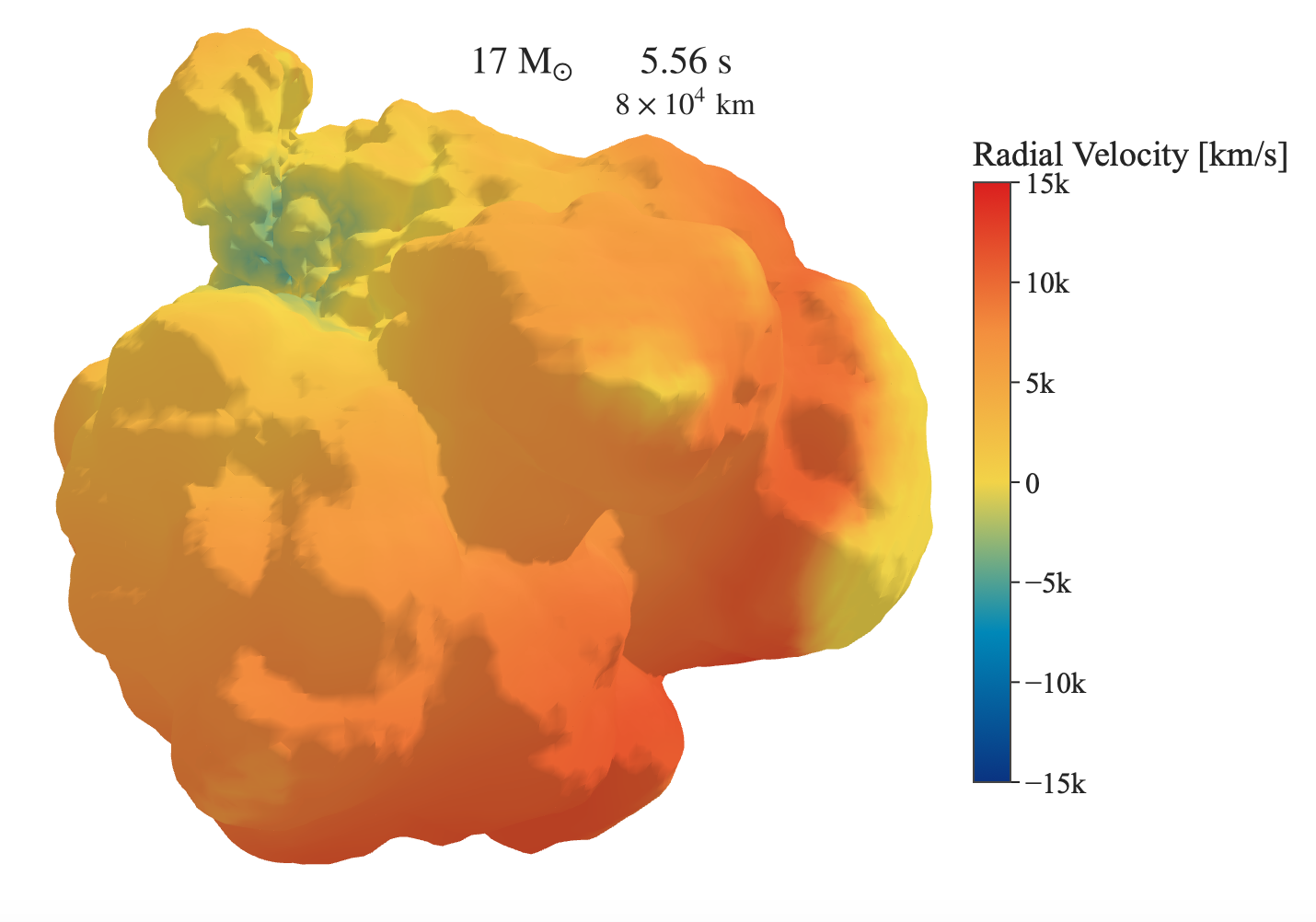}    
    \includegraphics[width=0.44\textwidth]{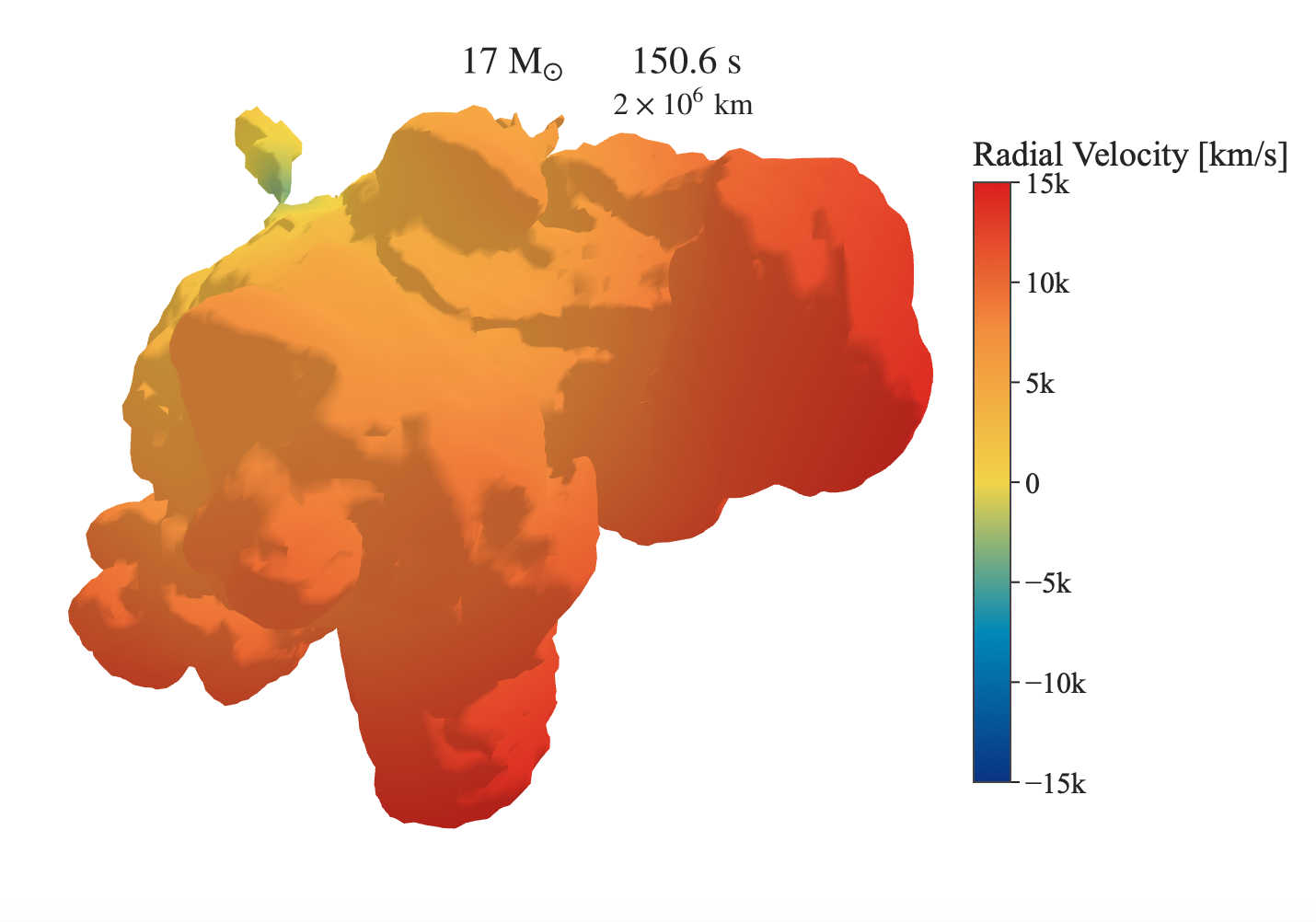}
    \includegraphics[width=0.44\textwidth]{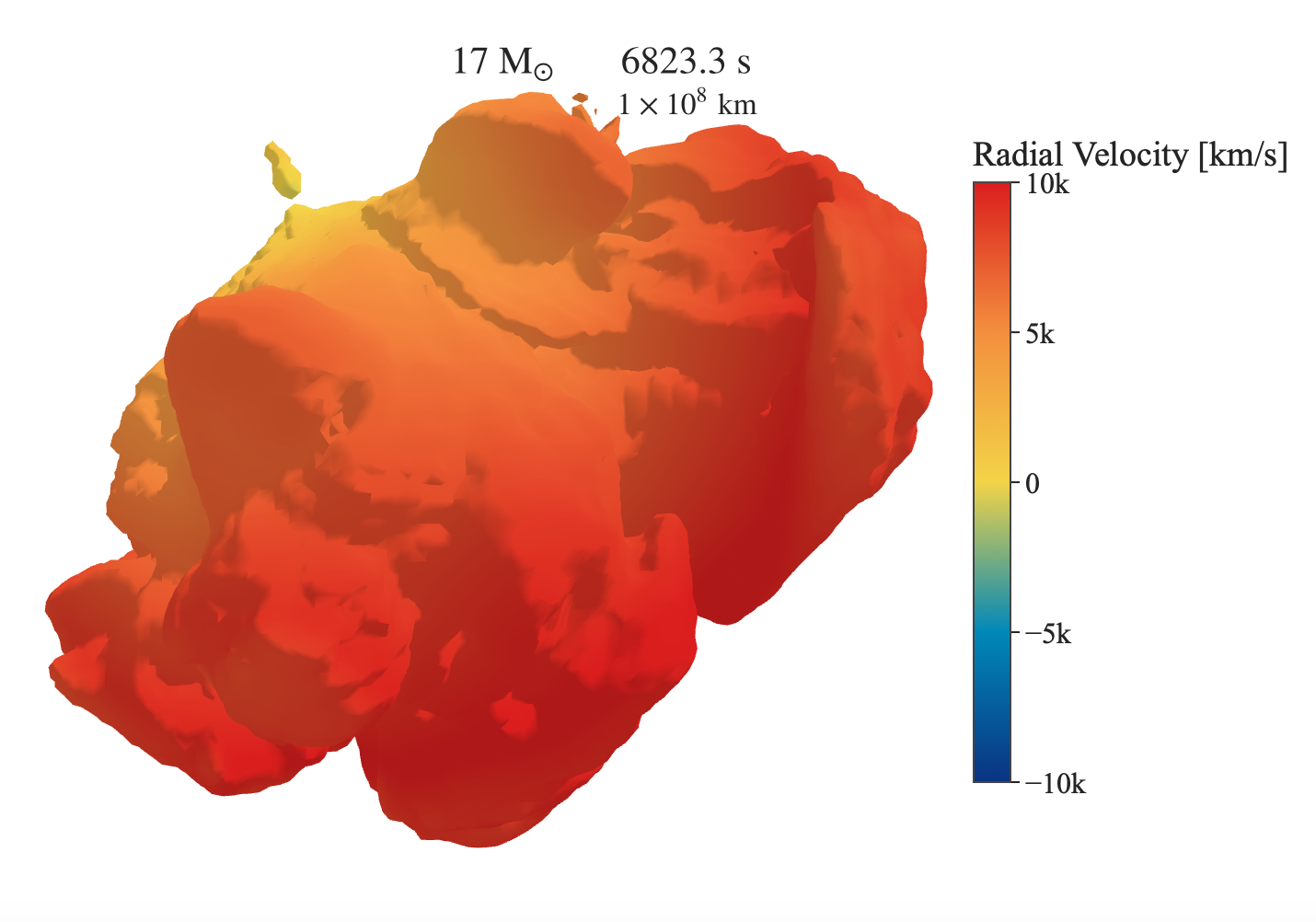}
    \includegraphics[width=0.44\textwidth]{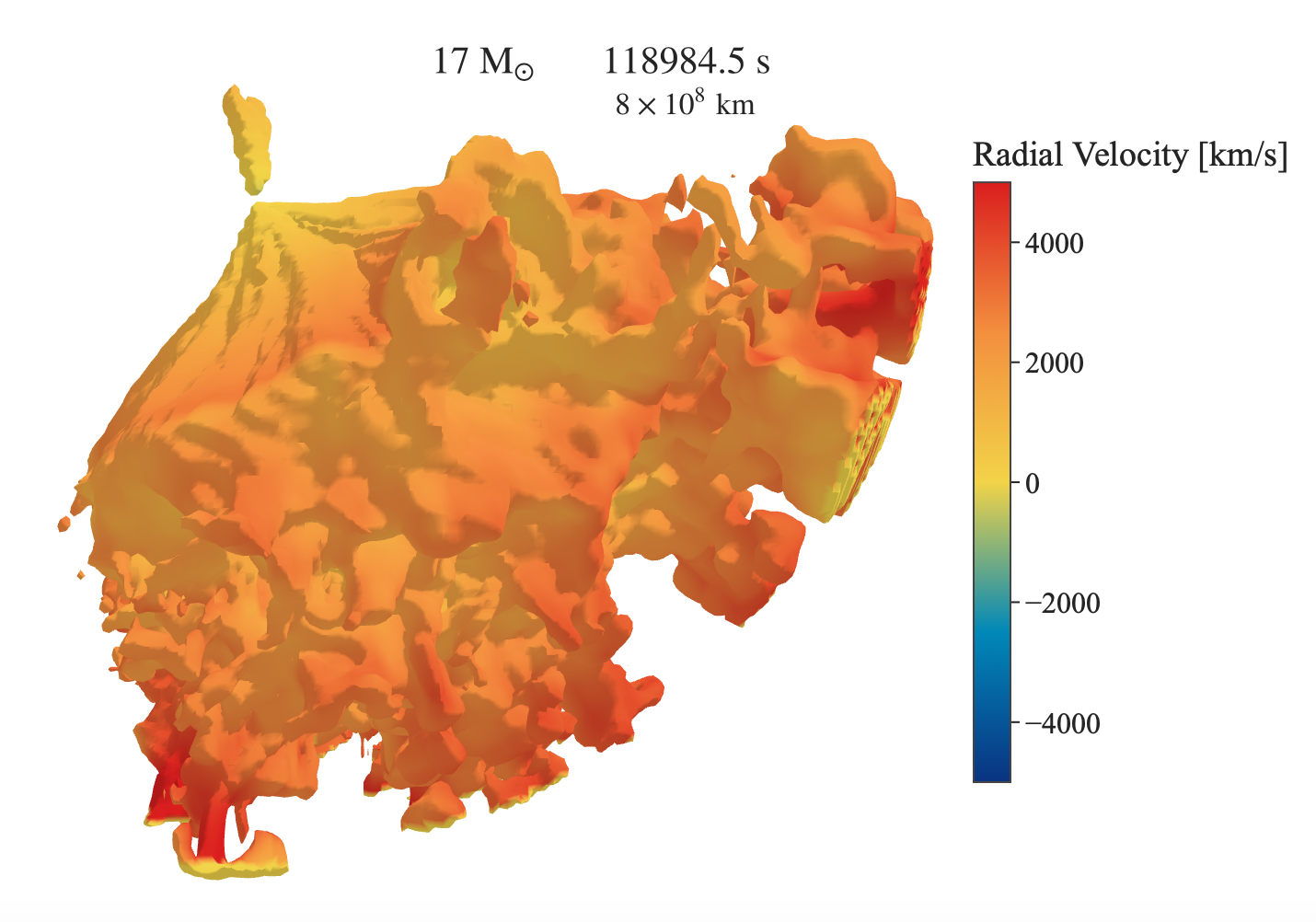}
    \caption{3\% mass fraction isosurfaces of the $^{56}$Ni distribution, colored by the radial velocities, at mapping from F{\sc{ornax}} to FLASH, at the intersection of the shock with the H/He interface ($\sim$146 s), interaction with the reverse shock ($\sim$6000 s), and after shock breakout. Note the different scales along the grid and in the color bar. The early geometry of the nickel, at seconds after core bounce, organically evolves into the structure at breakout, at more than a day after core bounce. The nickel ejecta remain predominantly in the southern hemisphere. In panel 2, the shock has just run into the interface. 1000s of later, in panel 3,  the bulk of the nickel begins to flatten upon interaction with the interface.  In panel 4, we see fragmentation of the outermost nickel, much of which is moving at velocities greater than 4000 km s$^{-1}$.} 
    \label{fig:nivel}
\end{figure*}

Fig.\,\ref{fig:ni_over} shows the nickel expansion within the metal core at $\sim$427\,s, before it interacts with the reverse shock. A large-scale, low-density  cavity is visible in the oxygen distribution, with the lower-density interior excavated by the ejection of a dominant nickel bubble surrounded by a higher-density shell. The core of the plume is composed primarily of nickel, which is surrounded by a sheath of silicon-, oxygen-, and helium-rich material (see also \citealt{sandoval}). The RTI triggered by the reverse shock partially shreds the plume surface into smaller scale filaments.

We visualize isosurfaces of the 3D $^{56}$Ni mass fraction distribution at various time snapshots from the onset of explosion to breakout in Fig.\,\ref{fig:nivel}.
The initial $^{56}$Ni distribution from the F{\sc{ornax}} model is highly asymmetric, with a subdominant plume in the northwest direction direction and a dominant multi-lobed plume in the southeast direction. When the reverse shock propagates back into the nickel-rich region, the most extended nickel plumes flatten and are compressed. Thereafter, we see marbling of the large-scale plume structure, as RTI tendrils of nickel punch through the H/He interface. RTI triggered upon interaction of the nickel plumes (just trailing the forward shock) with the reverse shock produce the familiar nickel `fingers', carved out by heavier, `ballistic' nickel parcels which move inertially into the reverse shock. We see just one dominant nickel structure in our model, whereas Cas A has three such structures, reflective of a tripolar explosion asymmetry. Such asymmetries can arise stochastically in the first seconds of explosion and persist, as we show here.

\begin{figure*}
    \centering
    \includegraphics[width=0.95\textwidth]{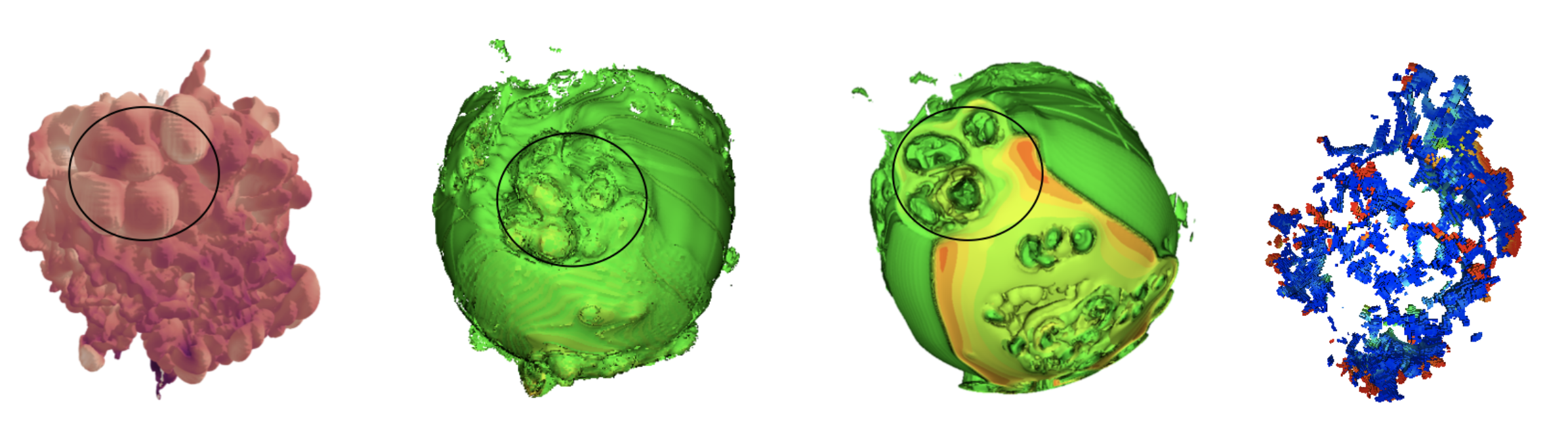}
    \caption{An isosurface of the $^{56}$Ni ejecta  (\textbf{left}) at $\sim$118,000\,s, after shock breakout, as seen along the southern hemisphere where breakout first occurs and the majority of the nickel lies; the shock surface (\textbf{second figure}) defined by a entropy isosurface of 10$^{11}$\,erg\,K$^{-1}$ at $\sim$85,000\,s, just before breakout; and (\textbf{third figure}) the shock surface identically defined after breakout at $\sim$120,000\,s. The wingspan is 8$\times$10$^8$\,km. We see `pinched fingers' of four nickel clumps (black circle) encompassing $\sim$0.001\,M$_{\odot}$ and moving at $\sim$4500\,km\,s$^{-1}$, similar to the nickel clump properties used to explain the Bochum event in SN1987A \protect\citep{1995A&A...295..129U}. Just before breakout, the nickel is just interior to the maximum shock surface at breakout, with the aforementioned clumps visibly deforming the shock surface on small scales. After breakout, as the shock leaves the simulation domain, we see in cross section the tunnels carved out by nickel by the same four clumps, leaving a clear ring-like structure, as well as by nickel clumps throughout. \textbf{Right:} Silicon (blue) and nickel (red) isosurfaces exterior to the reverse shock at $\sim$10\% of the respective maximum clump densities. We see the formation of silicon rings and nickel RTI decorating the rings into  crown-like structures. Such features have been identified in Cas A (see, e.g., Fig. 8 of \protect\citealt{orlando-cent}).}
    \label{fig:vel_ni}
\end{figure*}

One such feature is the development of four clumps of nickel in proximity along the surface of a large-scale plume, appearing as `pinched fingers' circled in Fig.\,\ref{fig:vel_ni}, left panel, showing an isosurface plot of the $^{56}$Ni and shock surface near breakout.  The nickel ejecta span $\sim$8$\times$10$^8$ km and is moving at an average velocity of $\sim$3400 km s$^{-1}$. The outermost $\sim$0.001\,M$_{\odot}$ of nickel in these clumps move at velocities of $\sim$4500\,km\,s$^{-1}$, similar to the characteristics of the nickel bullets invoked to explain the Bochum event in SN1987A in the first month of observation. \citep{1995A&A...295..129U}. By $\sim$20,000\,s, this cluster of four nickel clumps has caught up with, and deformed, the shock front. By $\sim$92,000\,s, the shock breaks out first along the southern hemisphere and we see nickel bullets not far behind, tunneling through the stellar envelope and leaving holes as exit wounds. After the first plumes leave our simulation grid, we are able to see the nickel trajectories in cross section. In addition, we are able to see, along a density-limited isosurface (e.g., \citealt{gabler,orlando-mil}), silicon `rings' punctuated by nickel RTI fragments to form crown-like structures. Note the similarity of both the nickels fingers and crown-like features to similar structures in Cas-A (e.g., \citealt{orlando-mil}). 

At the moment of breakout, the expanded nickel bubbles occupy $\sim$10$\%$ of the total star volume, but more than $\sim$30$\%$ of the total star surface area because of the clumpy structure. 
While the large scale structure is driven by the initial morphology of the neutrino-driven explosion, the small-scale `bubbling' is entirely a manifestation of nickel clumps deforming the shock surface. Comparing the nickel morphology over more than 100,000 seconds,
we find that the large-scale structure at breakout is strikingly similar to that at the initial mapping, but with structural details and additional small-scale features imprinted by the reverse shock and RTI.
This similarity across spatial and temporal scales extends to the shock surface as well, with earliest shock breakout occurring in the same direction as the dominant, most-extended neutrino-driven plumes in the first seconds of explosion. The overwhelming majority of $^{56}$Ni is ejected along the southern hemisphere, along the direction of the earliest shock breakout, with only $\sim$0.0004 M$_{\odot}$ ejected along the northern direction. We note that nickel clump and bullet evolution will differ by progenitor mass, sensitive to the explosion energy, the stellar envelope, and the reverse shock development. 

Expansion of the nickel bubble through the cavity, and possible compression upon interaction with the reverse shock, will leave behind a ring-like structure not dissimilar from the those observed in Cas A \citep{orlando-mil,2024ApJ...965L..27M}. At breakout, the dominant bulk of high mass fraction oxygen is ejected orthogonal to the nickel along two lobes, with a smaller lobe in the northern hemisphere. The silicon evolution follows closely, but lies interior to, the nickel structure. The titanium isotopes ($^{42}$Ti and $^{44}$Ti) have the countervailing pattern, following the nickel distribution along its outermost extent. These structures mimic in large part the distribution at the time of mapping. It would be of particular interest to see how these morphologies evolve out to breakout evolve on larger timescales, for centuries and through interaction with the ISM. For instance, observations of Cas A indicate large voids in silicon, iron, and titanium \citep{orlando-cent} seeded upon interaction with the reverse shock.

\subsection{Mixing}

\begin{figure*}
    \centering
    \includegraphics[width=0.47\textwidth]{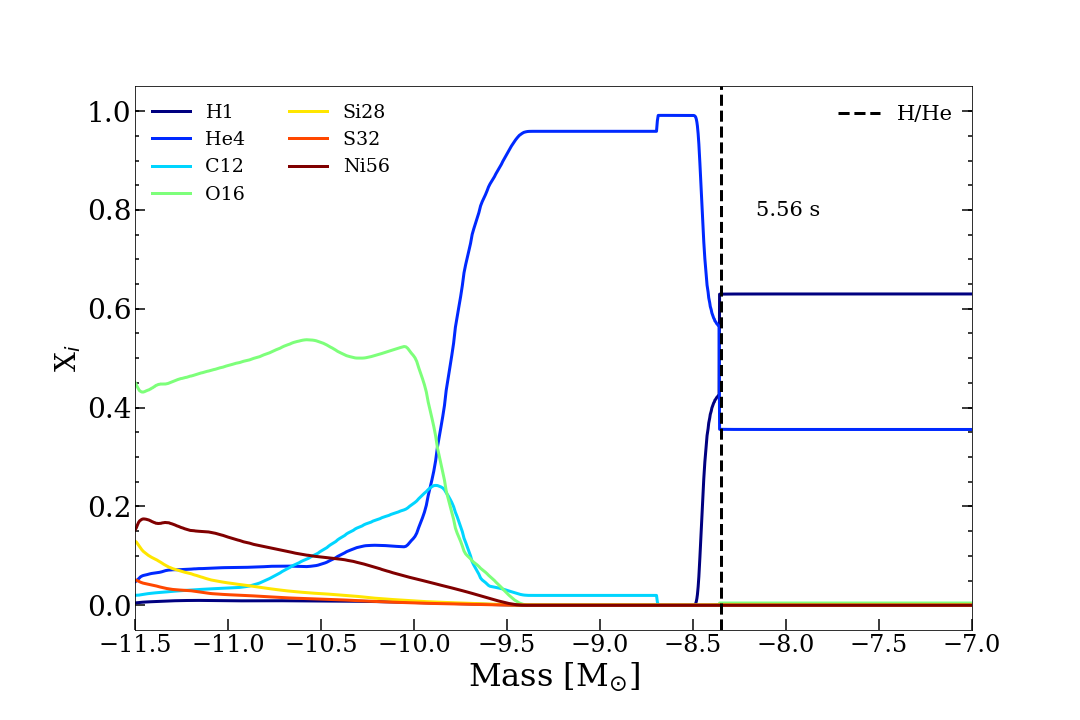} 
    \includegraphics[width=0.47\textwidth]{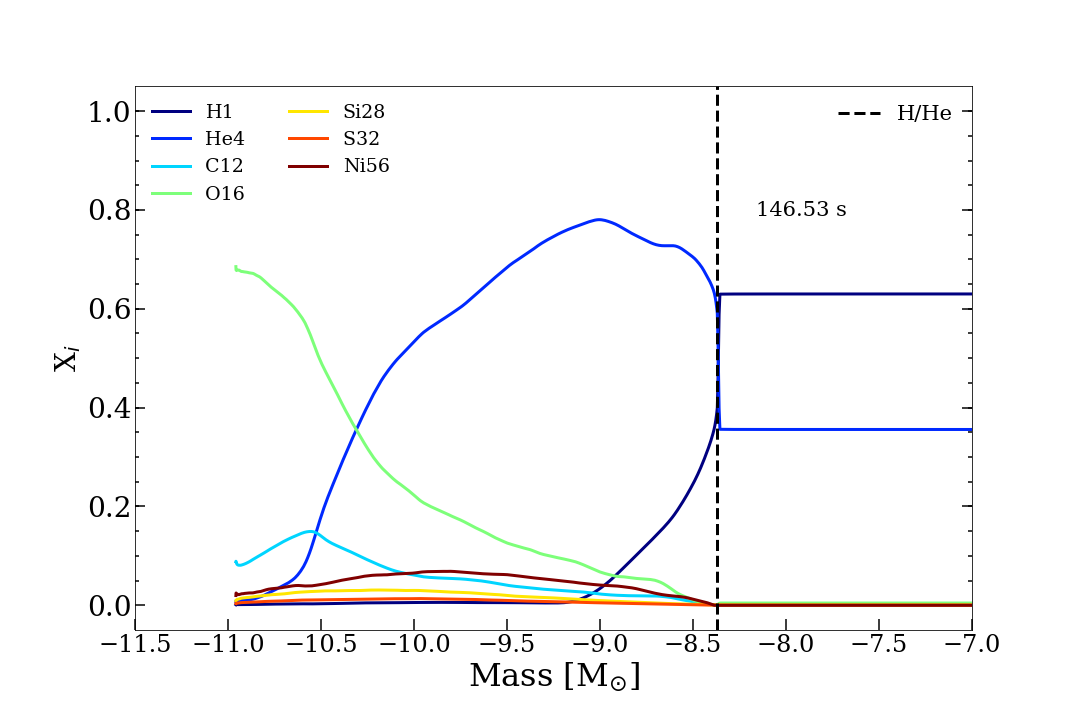} 
    \includegraphics[width=0.47\textwidth]{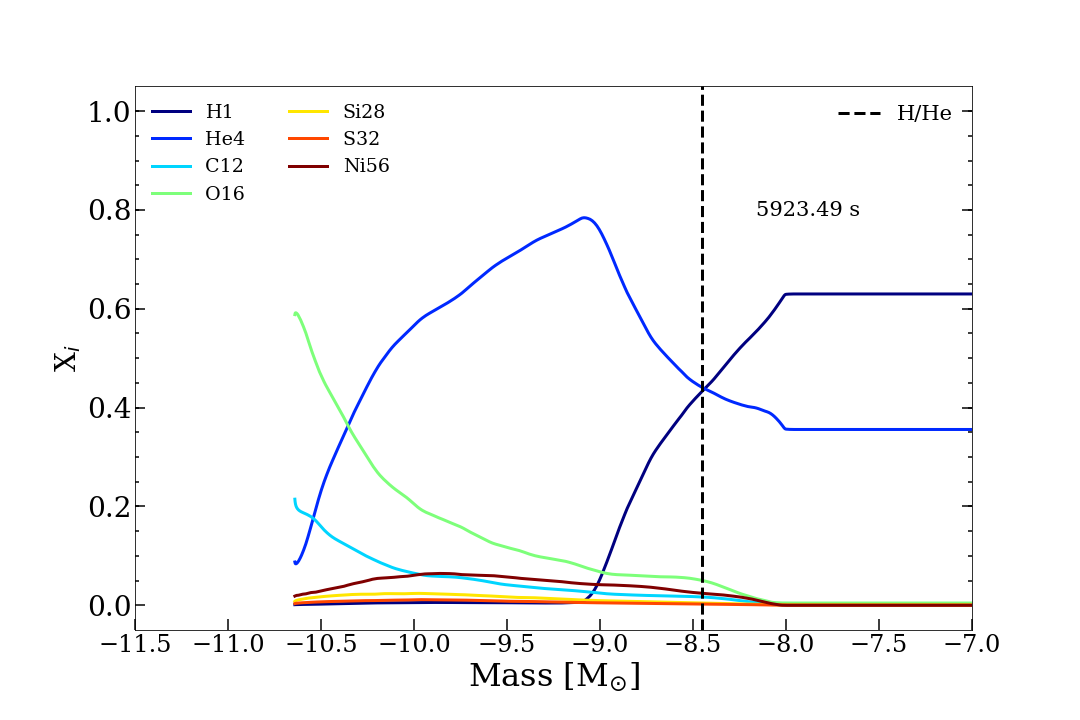} 
    \includegraphics[width=0.47\textwidth]{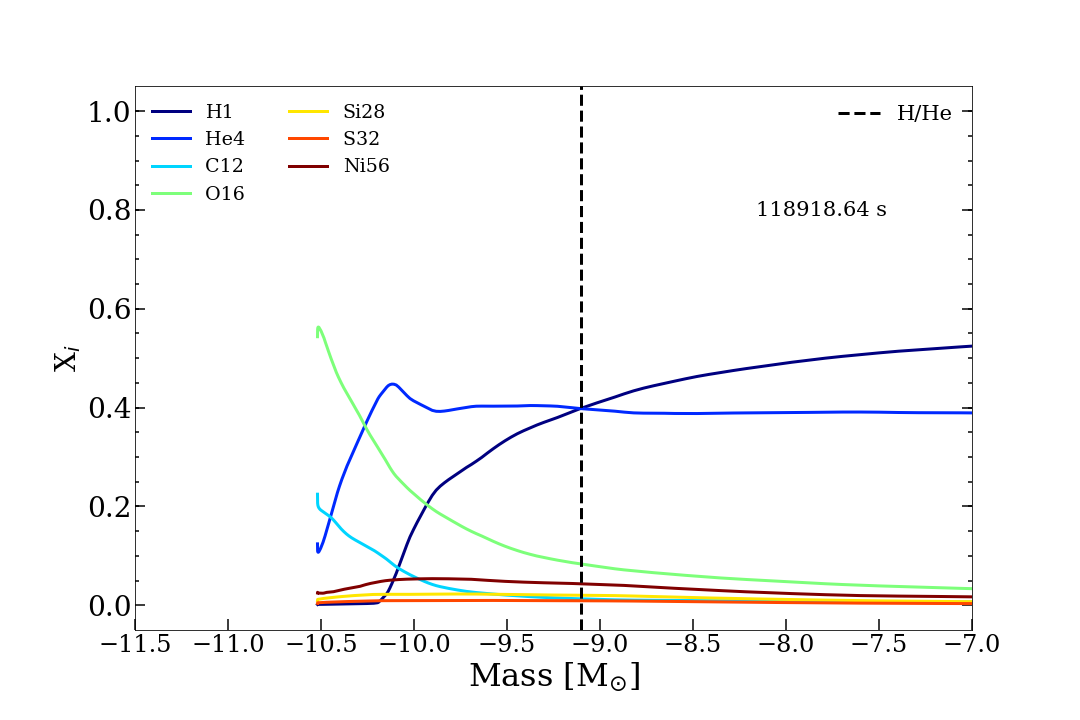} 
    \caption{Mass fractions of various isotopes vs depth from the stellar surface (in M$_{\odot}$) at the same four distinct times as in Fig.\,\ref{fig:dens_avg}. Significant mixing occurs in the bottom two panels as the metal core interacts with reverse shock at H/He interface. The H/He interface moves in by $\sim$0.7 M$_{\odot}$ due to mixing of $^4$He and metals outward, and H inward.} 
    \label{fig:rho_iso}
\end{figure*}

Mixing, calculated as the transport of elements across the angle-averaged H/He interface, begins upon interaction of the inner metal core with this interface, mediated by the reverse shock. RTI facilitates mixing of metals outward and hydrogen inward. To illustrate composition evolution, we show the angle-averaged chemical composition for our usual time sequence plotted against mass coordinate from the stellar surface in Fig.\,\ref{fig:rho_iso}. Together with Fig.\,\ref{fig:rho_init}, which shows the composition at core bounce, we illustrate the evolution of select isotope distributions in entirety, from core bounce to shock breakout. Mixing can be visually extracted by looking at the smearing of elements across the H/He interface. 

The majority of metals experience significant mixing ($>$60$\%$) outward into the hydrogen envelope, except for $^{12}$C, $^{16}$O, and $^{20}$Ne, which constitute the inner metal core and are mixed outwards by $\sim$35, 43, and 49$\%$, respectively. At breakout, more than 85$\%$ of the nickel has been mixed outward (Fig.\,\ref{fig:bullet}). An additional $\sim$0.9\,M$_{\odot}$, swept through by the reverse shock, has been mixed outwards by one day. This includes $\sim$0.6\,M$_{\odot}$ of $^{4}$He and the remaining $\sim$0.3\,M$_{\odot}$ in metals, predominantly nickel, oxygen, and carbon. Simultaneously, $\sim$0.2\,M$_{\odot}$ H is mixed inwards at breakout. This accounts for the H/He interface receding by $\sim$0.7 M$_{\odot}$ after breakout, seen in Fig.\,\ref{fig:rho_iso}. Motivated by measurements of nebular phase line profiles of hydrogen, mixed inward to velocities $<$700 km s$^{-1}$ \citep{utrobin}, we calculate corresponding quantities. Only $\sim$0.001\,M$_{\odot}$ of hydrogen is moving at velocities below 600\,km\,s$^{-1}$ at breakout. We follow the reverse shock evolution on the grid to find that after one week, $\sim$0.3 M$_{\odot}$ of H mixed inward, with the majority of it moving with mean velocities of $\sim$450\,km\,s$^{-1}$,  due to the later reverse shock formation in a massive progenitor. When comparing to \citealt{sandoval}, we  do not see a reverse shock forming at the C+O/He interface, largely because of the density profile differences set by the progenitor structure (compare our Fig.\,\ref{fig:dens_avg}, left panel, to their Fig. 3). We emphasize that our study through breakout, while neither a SN1987a or Cas A progenitor, nor carried out yet to the nebular phase, communicates a magnitude of effects through self-consistent 3D evolution similar to the spread of observed properties. 

At mapping, the cells with nickel mass fractions greater than 1$\%$ contain $\sim$10$\%$ of the kinetic energy (though less than 1$\%$ of the total ejecta mass), dropping to $\sim$5$\%$ at breakout.  If we add also the isotopes of elements comoving with the nickel, we find that cells with nickel ejecta transport $\sim$40$\%$ of the kinetic energy at the time of mapping. As the shock goes down the density gradient and accelerates, a smaller fraction of kinetic energy, $\sim$23$\%$, remains in the metal ejecta comoving with the nickel (again, despite constituting $<$10$\%$, $\sim$0.8 M$_{\odot}$ of the ejecta mass). Once the shock decelerates at the the H/He interface at $\sim$146\,s, the nickel ejecta assume a greater fraction of the kinetic energy (only by several percent) until colliding with the reverse shock. Because of the delayed 
reverse shock, nickel accelerates out longer, to $\sim$6000\,seconds.  At $\sim$10,000\,seconds, the fractional kinetic energy in nickel decreases due to the reverse shock and this continues until breakout. After breakout, the shock again accelerates down the density gradient into the CSM, and the nickel ejecta will lag further behind. 

Thorough nickel mixing into the stellar envelope,  the large-scale deformation of the aspherical aspherical shock front with fast moving nickel clumps, and a lengthy delay in shock breakout time by direction can lead to an early rise in polarization, as perhaps seen in SN2012aw, 2013ej and SN2023ixf (though concurrent flash ionization features for the latter perhaps prefer an asymmetric CSM as the origin of the early polarization), before the helium and metal core is exposed and absent the need of an asymmetric CSM \citep{2024A&A...687L..17N,2024ApJ...975..132S}. The early steep rise in the SN2023ixf light curve may be explained by such an aspherical shock breakout \citep{2024arXiv241019939K}.

We emphasize here our study of a RSG progenitor. Absent, or with a lighter, hydrogen envelope, we expect higher velocities unmitigated by a reverse shock. Additionally, expansion into a low-density CSM will allow the fastest nickel bullets to further accelerate until cooling by radioactive decay of $^{56}$Ni $>$  $^{56}$Co $>$ $^{56}$Fe in the optically-thin CSM saps their energy, on timescales of days to weeks. The bulk of the ejecta trailing interior will remain optically thick for longer and sustain longer acceleration, including via radioactive decay. Competing timescales, namely the time of reverse shock formation, which fragments the nickel ejecta, and the half-life for the decay chain will in concert shape the final structure and dynamics of the metal ejecta. The former is entirely due to the energetics and progenitor profile of the early CCSN explosion. We will explore the development of breakout morphology by progenitor mass in an a subsequent paper (Vartanyan et al., in prep.). 

\section{Post-Breakout Evolution into the CSM}\label{sec:csm}

We continue our simulations into the CSM to follow the evolution post-breakout, mapping at a time of $\sim$118,000 s post-bounce. In this preliminary look, we use spherically-symmetric toy models for a CSM powered by stellar winds with two variations. Our CSM density follows a normalized  r$^{-2}$ power-law, which encompasses $\sim$0.01 M$_{\odot}$ of material in two different radial distributions, one extending to $\sim$2.4$\times$10$^{14}$ cm and the other to $\sim$1.1$\times$10$^{15}$ cm. We continue these simulations to an additional $\sim$5.6 and $\sim$1.4 days, respectively. The structure of the CSM can critically alter the emergent diagnostics \citep{2024arXiv240219005C,2023A&A...677A.105D,2024arXiv240402566T} and merits a systematic 3D study. That is not our intent here. 

\begin{figure*}
    \centering
    \includegraphics[width=0.47\textwidth]{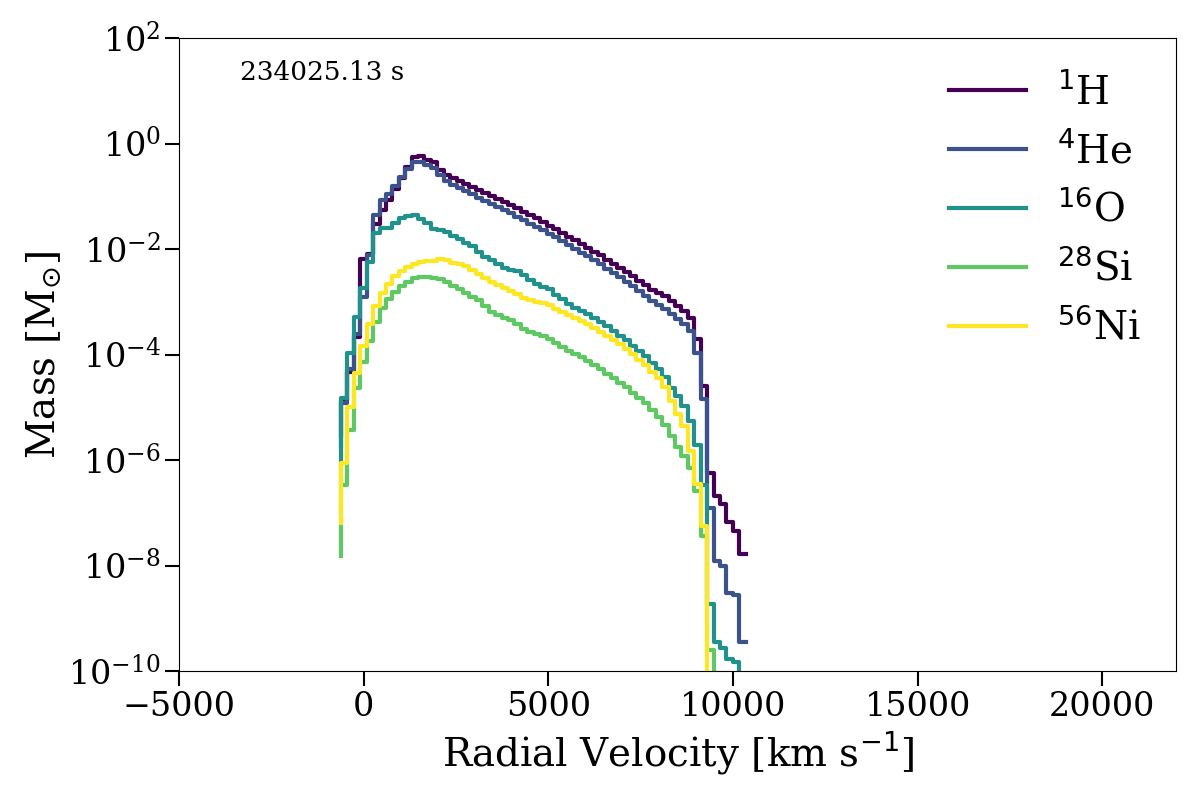}
    \includegraphics[width=0.47\textwidth]{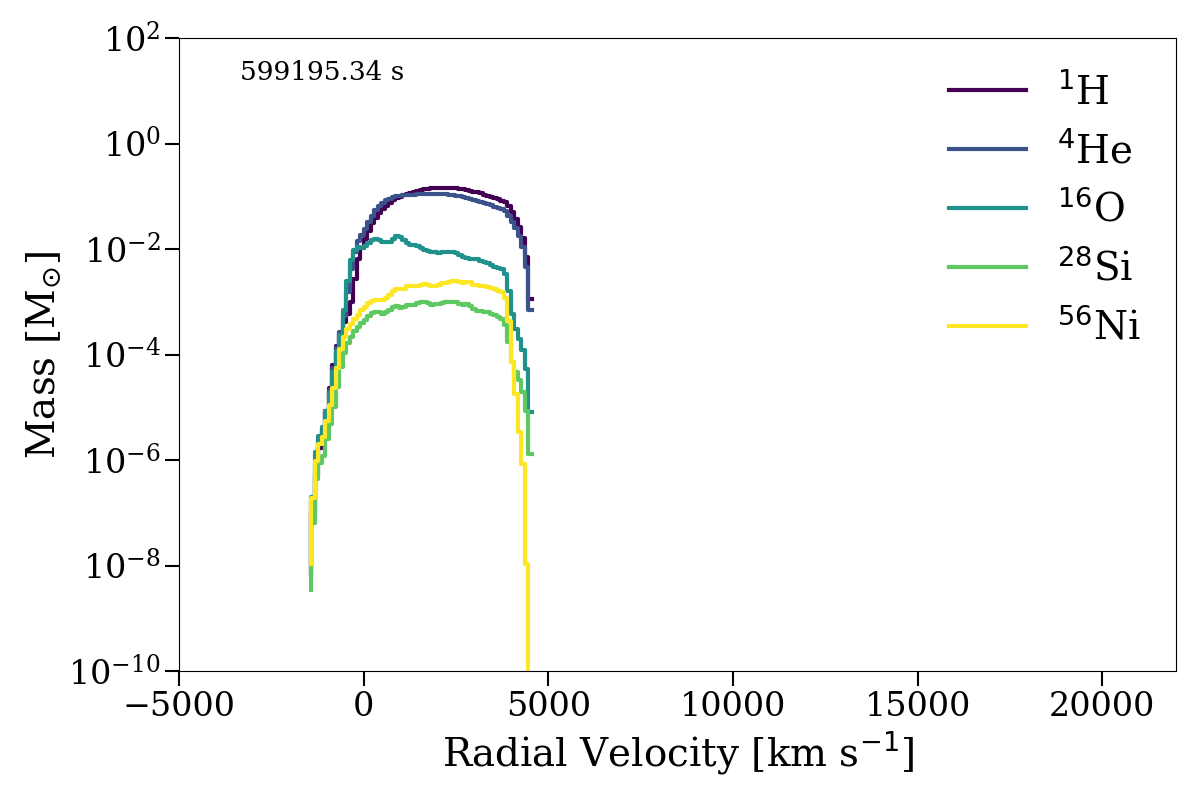}
    \includegraphics[width=0.47\textwidth]{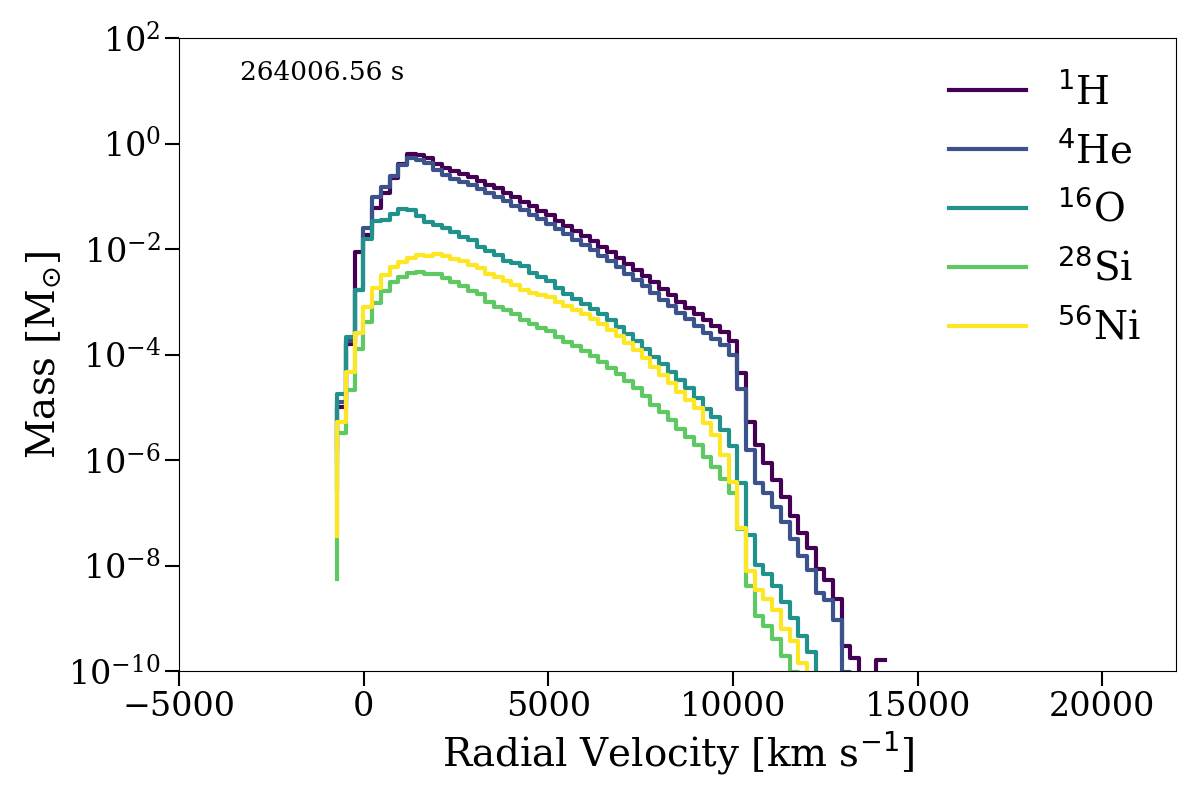}
    \caption{Same as Fig.\,\ref{fig:iso_hist}, but now following the isotope velocities after shock breakout into the CSM. The top panel shows the higher density CSM configuration. Initially, the matter accelerates into the CSM, with the highest velocity nickel reaching velocities of $\sim$10,000 km s$^{-1}$. By $\sim$234,000 s post-bounce, enough matter has been swept up to decelerate the nickel. \textbf{By $\sim$599,000 s, $\sim$20$\%$ of the nickel has left the outer simulation domain, with the remaining decelerated below 5000 km s$^{-1}$,  slower than at shock breakout.} The bottom panel illustrate the lower density CSM configuration. At  $\sim$264,000 s post-bounce, the nickel is still accelerating, reaching velocities of $\sim$13,000 km s$^{-1}$. }
    \label{fig:iso_hist_CSM}
\end{figure*}

Fig.\,\ref{fig:iso_hist_CSM} shows histograms for various isotope mass fractions as a function of radial velocity (km s$^{-1}$) at different times, akin to Fig.\,\ref{fig:iso_hist}, but now for the evolution into the CSM. The top panel shows the higher density configuration. The explosion accelerates into the CSM down the density gradient, with the fastest nickel parcels reaching velocities of $\sim$10,000 km s$^{-1}$. By $\sim$115,000\,seconds post-mapping, enough matter has been swept up to decelerate the nickel. \textbf{
By $\sim$600,000 seconds, $\sim$0.02 M$_{\odot}$ of nickel, with a sufficient acceleration history, has left the outer simulation domain, and the remaining $\sim$0.08 M$_{\odot}$ of nickel moves at velocities below 5000 km s$^{-1}$, slower than at shock breakout.} The first exodus of nickel from the grid begins by $\sim$290,000 seconds. We note, however, that the nickel decelerates well before.The bottom panel illustrates the lower density CSM configuration. At  $\sim$115,000\,seconds post-mapping, the nickel is still accelerating, reaching velocities of $\sim$13,000\,km\,s$^{-1}$. We emphasize that these estimates neglect radiative cooling. The most extended nickel structures with the fastest velocities in low-density, optically thin, regimes would lose energy through nickel-chain decay and result in slower velocities than our estimate in the CSM. The final dynamics depend on the distribution of nickel and the ambient densities. We will investigate this in future work. 


\section{Observational Predictions}\label{sec:sedona}

\begin{figure*}
    \centering
    \includegraphics[width=4in]{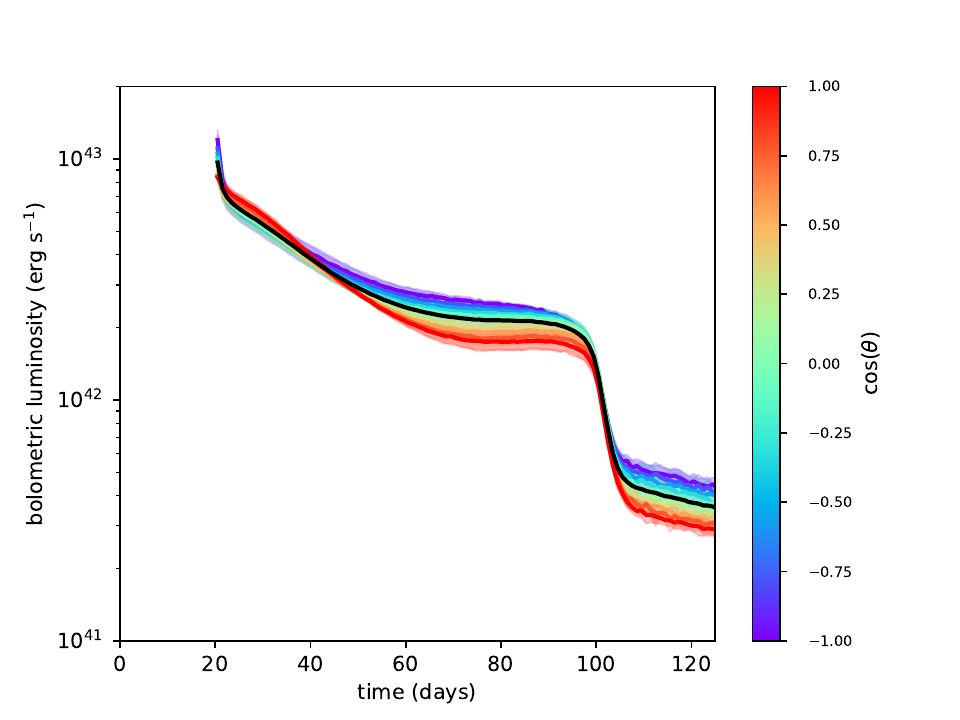}
    \caption{Bolometric light curve of the model. The thick black line shows the angle-averaged values, while the thin colored lines show light curve from various viewing angles. On the plateau, the bolometric luminosity varies by a factor of $\sim 2$ depending on the line of sight.}
    \label{fig:bolometric_lc}
\end{figure*}

We have calculated the observable properties of the  model (based on the second, lower-density CSM configuration discussed above) using 3D radiative transfer calculations that cover the phases 20 days to 120 days. Fig.~\ref{fig:bolometric_lc} shows the synthetic bolometric light curve over these epochs and from multiple lines of sight. The light curve morphology resembles that of a standard Type-IIP supernova, with a extended plateau of slowly declining luminosity, followed by a rapid drop-off to an exponentially declining, radioactively powered tail. The luminosity on the plateau is $\sim$2$ \times\,10^{42}$\,erg, comparable to that of moderately bright SNeIIP such as SN2004et \citep{maguire2009}. The duration of the plateau is around 100 days, comparable to that of SN2004et, and within the range of 80 $-$ 120 days found among the sample of SNe~IIP.

\begin{figure*}
    \centering
    \includegraphics[width=6in]{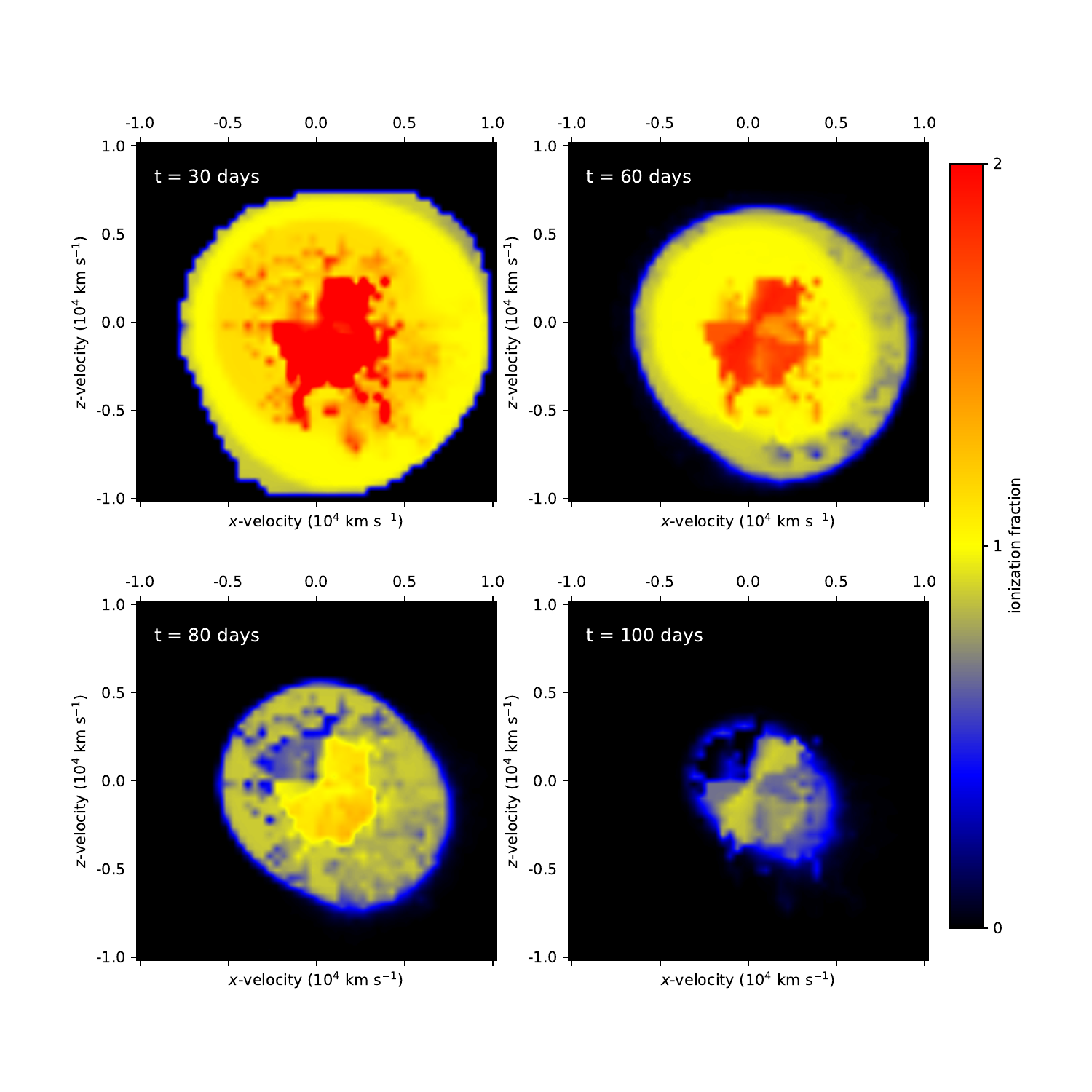}
    \caption{The evolution of the ionization state of the ejecta shown in a slice through the x-z plane. The ionization fraction (defined as the electron number density divided by total number density) reaches a maximum value of 1 in a regions of fully ionized hydrogen and higher values in regions of multiply ionized heavy elements. Over time, as the ejecta expand and cool, the ionization front recedes in velocity coordinates, with the ejecta becoming largely neutral near the end of the plateau ($\approx 100$~days).  The electron scattering photosphere is nearly coincident with the ionization front and shows a bulk asphericity, contributing to the anisotropic luminosity of the supernova. }
    \label{fig:ionization_state}
\end{figure*}

During the plateau phase, thermal radiation energy deposited in the ejecta by the explosion shockwave diffuses out gradually. The hot inner parts of the hydrogen layers are ionized and a sharp recombination front forms around the LTE recombination temperature of hydrogen, $T_{\rm rec} \approx 6000$~K. Fig.~\ref{fig:ionization_state} illustrates the evolution of the ionization state over time. As the ejecta expand and cool, hydrogen recombines and the front moves inward in the comoving velocity coordinates. As electron scattering dominates opacity, the photosphere nearly coincides with the recombination front. The regulation of the photosphere temperature near $T_{\rm rec}$ leads to the slowly evolving luminosity on the plateau.

The observed bolometric luminosity  varies with viewing angle by as much as a factor of $\sim$1.6, being brighter at most phases from orientations where the nickel plume is moving towards the observer. 
The anistropic luminosity on the plateau reflects the asphericity of the supernova photosphere.
Because the flux at the photosphere should be roughly blackbody at the nearly fixed recombination temperature $T_{\rm rec}$, the luminosity observed from a certain orientation should be approximately proportional to the projected area of the emitting surface along that line of sight \citep{Darbha21}. Fig.~\ref{fig:ionization_state} shows that at day~80 the photosphere is roughly ellipsoidal with an axis ratio $\approx 1.25$, consistent with the factor of $1.6$ variations in luminosity with viewing angle at these times.  Doppler boosting  can also produce anisotropic emission, but given the average ejecta speeds in the model, $v \approx 5000$~km~s$^{-1}$, the  boosting effects are only at the $\approx 10\%$ level.  

The end of the light curve plateau occurs when the recombination front has receded nearly completely through the hydrogen layers and passes quickly through helium rich regions. The bulk of the ejecta rapidly becomes neutral and transparent and nearly all of the residual explosion energy escapes. The luminosity then drops sharply to the level supplied by the continuous energy ejection from the radioactive decay of the $^{56}$Ni chain.  As this transition is global, the end of the plateau is observed at nearly the same time when the system is observed from any viewing angle. 

The hydrogen and helium regions are neutral and transparent after the plateau, but the ejecta rich in heavier species (silicon and iron group elements, which have lower ionization potentials) may remain modestly ionized for some time after. On the early light curve tail ($\sim 110-150$~days) the $^{56}$Ni regions remain moderately optically thick to a combination of electron scattering and blended bound-bound transitions. Given the asymmetric distribution of nickel, the luminosity on the radioactive tail continues to shows variations by a factor of $\sim 2$ with viewing angle.  Non-LTE effects  become increasingly important after the plateau which calls into question the reliability of our LTE calculations in capturing the optical depth and luminosity anisotropy at these phases.

\begin{figure*}
    \centering
    \includegraphics[width=4in]{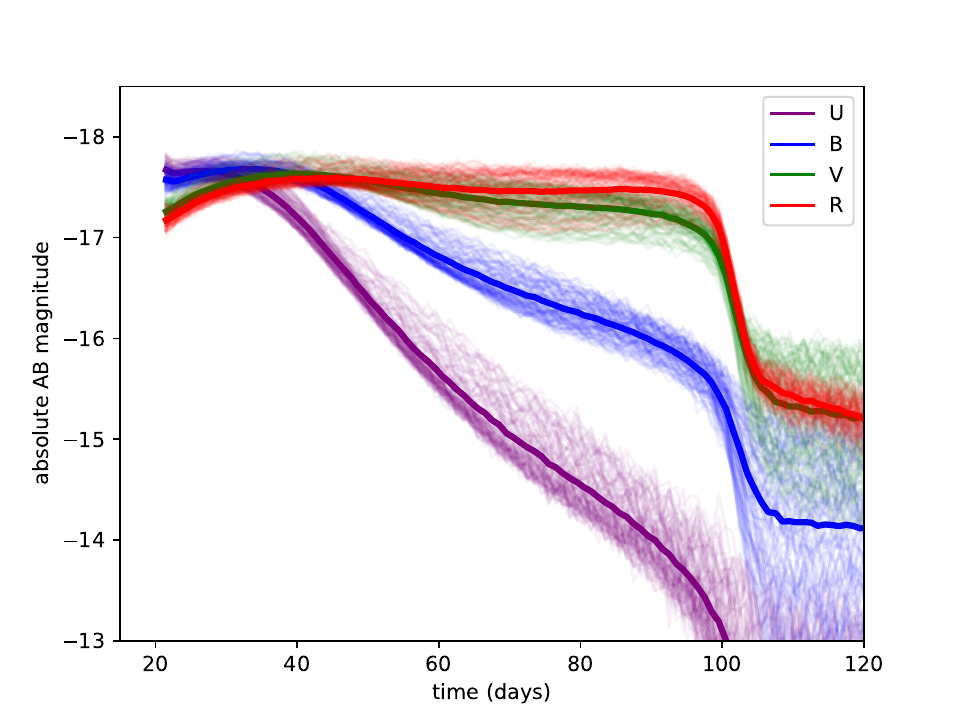}
    \caption{Broadband light curves of the model. The  thick lines shows the angle-averaged values, while the thin  lines show the spread in light curve from various viewing angles. Small scale fluctuations are due to numerical Monte Carlo noise. The general evolution towards redder colors at later times resembles that of observed Type II-P supernovae.}
    \label{fig:band_lc}
\end{figure*}

Fig.~\ref{fig:band_lc} shows the model synthetic broadband light curves from various viewing angles. The light curves qualitatively resemble those of normal SNe II-P, with a progressive evolution to redder colors. On the plateau, the $V-R$ color remains relatively constant, as the  photospheric temperature is regulated to be near the hydrogen recombination temperature. The U and B bands decline more sharply on the plateau due to increasing line blanketing from  numerous bound-bound transitions from iron-group species. The orientation dependence of the broadband light curves reflects the difference in projected photospheric surface area from different lines of sight.  The dispersion in broadband magnitudes increases significantly after the plateau, however given the neglect of non-LTE effects our calculations can not be expected to predict reliable colors as the ejecta transition to the nebular phase.

Such sharp variation in viewing angle by peak brightness, by as much as one magnitude, was noted in \cite{2024arXiv241020829M} for their ultra-stripped 3.5-M$_{\odot}$ progenitor, together with delays of days in the time of peak light. Though the authors attribute the variation to a small ejecta mass, we find similar variations in the light curve plateau for our model with a high ejecta mass. Likewise, \cite{2023ApJ...952..155Z} find variation in the plateau luminosity by a factor of $\sim$2 due to model variations in the stellar radius. We do not yet capture here variations in the light curve at peak, but asphericity of explosion will affect the light curves and spectra, and we dedicate thorough discussion to a future work.

\section{Conclusions}\label{sec:conclusions}


Before presenting our conclusions, we first broadly summarize limitations in CCSN study. CCSN observations suffer from gross systematic uncertainties \citep{2024arXiv241014027B} in estimating bulk explosion and progenitor properties, which still remain crude. As a recent illustration, estimates of  SN2023ixf progenitor mass vary from $\sim$12$-$18 M$_{\odot}$, and explosion energies from $\sim$1$-$3 Bethe \citep{2024A&A...681L..18B,2024MNRAS.tmp.2013Q,2024arXiv240600928M}, with nickel yields of $\sim$0.04$-$0.06 M$_{\odot}$.\footnote{Producing a robust explosion of several Bethe with a comparably small nickel mass is difficult, given strongly monotonic trends between the two in detailed CCSN simulations \citep{burrows2024}. Though it may be premature to draw conclusions here, a high explosion energy could suggest either additional contributions to the central mechanism, or sustained accretion and formation of a black hole as the remnant (\citealt{burrows2023,2024arXiv241207831B}.). The latter could naturally explain the dearth of observed nickel, with an appreciable fraction falling back onto the black hole.} 

Uncertainties in observations meet head-on with uncertainties in simulation and theory.
In key ways, CCSN theory is hostage to limitations in late-stage stellar evolution, including the development of convection \citep{2020MNRAS.493.4333R}, nuclear burning reactions \citep{farmer_variations_2016}, winds \citep{2017A&A...603A.118R},  convective overshooting \citep{2019MNRAS.484.3921D}, shells mergers and the transition from  convective to radiative nuclear burning \citep{sukhbold2018}, the role of mass resolution \citep{sukhbold2018} and nuclear network size \citep{farmer_variations_2016} and importantly, the importance of multidimensional progenitor models \citep{muller2017}, are now more than ever limiting factors in the successful advancement of the longstanding CCSN problem.

These limitations extend to extant breakout simulations. Supernova breakout studies to date $-$ including this study $-$ do not yet include multi-dimensional stellar progenitors that account for the onset of instabilities in the silicon- and oxygen-burning shells in the final moments before collapse. 3D simulations of oxygen-burning in the final stages of stellar evolution out to core collapse have only recently become available and only for a limited set of progenitors \citep{muller2016, fields2020,fields2021,2019MNRAS.484.3921D,2020ApJ...890...94Y}. Perturbations (e.g., \citealt{2013ApJ...778L...7C,2020ApJ...890...94Y,muller_janka_pert,burrows2018}) encourage successful CCSN explosion as seen in early simulations \citep{vartanyan2022,muller2019} and provide a more physically-motivated driver for seed asymmetries.

Equally absent are CCSN progenitor models with multi-dimensional envelopes. Convection in the stellar envelope, in addition to driving asymmetric pre-collapse mass loss and outbursts \citep{reilly} that can populate the CSM \citep{2022ApJ...936...28T}, also affects shock breakout dynamics and light-curve predictions \citep{goldberg2}. Degeneracies in progenitor radius, explosion energy, and ejecta mass complicate straightforward observational interpretation.

We note several limitations in our study. We continue shock breakout until homologous expansion, a prerequisite for subsequent follow-up with Sedona. Consequently, we do not yet calculate the light-curves at shock breakout, before homology is reached. As noted in \cite{sandoval}, a limitation is the use of the Helmholtz equation of state, which assumes all species are ionized. This only becomes important in the outermost cooler layers of the 17-M$_{\odot}$ progenitor, relevant in the moments of shock breakout. We mitigate this with an early mapping to Sedona. In this same interstitial regime,  during the days after breakout, we do not include nickel and cobalt radioactive decay heating in our FLASH evolution. The net energy contribution of $\sim$0.1 M$_{\odot}$ $^{56}$Ni is $\sim$2$\times$10$^{49}$ erg (only a fraction of which is deposited in the ejecta), small compared to our bulk kinetic energy $\sim$10$^{51}$ erg. However, radioactive heating would contribute to inflating nickel bubbles and affect the final morphology and dynamics \citep{gabler}. The significance of radioactive heating depends on the nickel abundance morphology.\footnote{For instance, ejected nickel may be shredded by a reverse shock that redistributes nickel on sufficiently short timescales from an initial local configuration to small pockets spread globally, as we see for lower mass progenitors (in prep.). This will affect energy deposition from decay accordingly. Forward and reverse shock dynamics are progenitor-dependent, and we will present a study of the complex shock morphology across diverse progenitor models in an upcoming study.} Additionally, we do not invoke in the early seconds of our shock propagation simulation a neutrino wind as an inner boundary condition (but see \citealt{2024arXiv241207831B}) nor continue nucleosynthetic burning. This is rather our advantage $-$ our CCSN simulations with F{\sc{ornax}} have continued significantly later in 3D with neutrino heating and nucleosynthesized yields (incorporated with SKYNET) than competing efforts, and so we are able to follow the detailed neutrino heating and CCSN nucleosynthesis for longer and its contribution at mapping is commensurately weaker. Regardless, additional efforts to refine the inner boundary condition, improve the equation of state transition at hydrogen recombination, and include heating by radioactive decay are planned.

Our work here presents an introduction to a systematic anthology of massive stellar explosions carried out to breakout and beyond. We summarize our key results below. 

\begin{itemize}

\item{We explode a 17-M$_{\odot}$ with an explosion energy of $\sim$1 Bethe and a nickel yield of $\sim$0.1 M$_{\odot}$ in 3D for core bounce to beyond shock breakout. Shock breakout is very asymmetric, first along the southern hemisphere at one day, then along the northern hemisphere a day later.}
\item{The morphology of the early explosion is well-preserved in time to beyond shock breakout and past $\sim$a week, the extent of our simulation.}
\item{We find velocities as high as $\sim$14,500 km s$^{-1}$ for the nickel ejecta, and shock velocities as high as $\sim$23,000 km s$^{-1}$, following the density profile of the envelope in their acceleration and deceleration.}
\item{The shock hits the hydrogen/helium interface at $\sim$146 seconds. At $\sim$6000 s, a reverse shock forms and propagates inward in mass, though it continues to move outward in radius. At equatorial breakout, $\sim$117,000 s, the reverse shock has begun to propagate inward in radius, plowing a width of $\sim$3$\times$10$^{8}$ km with the forward shock. RTI-triggered fingers penetrate through the reverse shock, in some cases spanning the entire shock width.}
\item{Near breakout, we see the fastest nickel ejecta tunnel through, but never quite surpass, the shock. We see four clumps of nickel, with a combined mass $\sim$1$\times$10$^{-3}$ M$_{\odot}$, in the southern hemisphere moving at velocities greater than $\sim$4500 km s$^{-1}$. Though we do not intend to directly compare results, and post-breakout velocities are sensitive to the  CSM structure, such a result is reminiscent of the Bochum event in 1987A. }
\item The nickel ejecta follow the entropy plumes entrained interior to the shock through the simulation, carving out cavities that persist as large-scale voids in the oxygen metal core, and the evolution of their respective distributions  is thus anti-correlated.
\item{Nearly the entirety of the nickel is mixed outward at breakout. Days afterward breakout, $\sim$0.3 M$_{\odot}$ hydrogen is mixed inward, with the majority moving at velocities below $\sim$600 km s$^{-1}$. Post-breakout follow-up is necessary to follow the reverse shock and mixing inwards for massive progenitors. SN1987A, with a possible BSG or binary progenitor, has perhaps several solar masses of hydrogen mixed inward to these velocities in the nebular phase \citep{utrobin}.}
\item{At breakout, the bulk of the nickel is moving at mean radial velocities of $\sim$3400 km s$^{-1}$. }
\item Our predicted light curves are consistent with a moderately bright Type II-P CCSNe, with a plateau and radioactive tail luminosity varying by a factor $\sim$2 by viewing angle.

\end{itemize} Initial asymmetries set by neutrino-driven convection and the many-second interplay between explosion and accretion result in large-scale asphericities, with further modification at the H/He interface and emerging RTI, at shock breakout. 
In our simulated second-to-day timescales, neutrino-heated bubbles of nickel, trailing behind the shock, preserve their structure. These features will likely be frozen-in to much longer timescales \citep{gabler,orlando-cent,orlando-mil}. Fast-moving nickel bullets can escape significant deceleration by the reverse shock and penetrate towards the forward-moving shock.  Higher energies  ($\sim$1 Bethe) in CCSN simulations correlate with greater asymmetries \citep{burrows2024}. Whether our energetic and highly asymmetric 17-M$_{\odot}$ model presents a canonical type II-P supernovae event, and whether we expect such large asymmetries in observations, remains a fruitful pursuit.

Our intent here is not to replicate Cas A, SN1987A (which have very different envelope structures than the RSG we study here), or any observed CCSN, but rather to illustrate that an energetic massive star CCSN followed in detail from late-time neutrino evolution to shock breakout $-$ as a model for a massive star progenitor $-$ can produce many observed properties of CCSNe as a natural consequence. We argue that the emergent asymmetries, including efficiently-mixed high velocity asymmetric ejecta, are generic results of massive progenitors ($\gtrsim$15-M$_{\odot}$) undergoing energetic CCSN and carried out to breakout. Low-mass stars ($\sim$8$-$13 M$_{\odot}$), which explode earlier with a lighter mantle, will rather evince smaller-scale asymmetries, bundled with weaker explosion energies and lighter nucleosynthesis. 

Late-time simulation is essential not only for the CCSN context, but also for the study of post-breakout morphology and diagnostics. Synthesizing these two components self-consistently is necessary to couple the early instabilities driven by neutrino heating to the correlated structures days later. Though the large- and small-scale asphericity of simulated and observed CCSNe has long been known, we emphasize here the degree to which this asymmetry, perhaps overlooked, matters. This renewed perspective will frame the study of breakout structures and times, provide insight into the long-term sustained fallback onto the remnant, and inform the interpretations of the critical angle-dependence in observed line profiles, electromagnetic signatures, and ejecta morphologies.

\section*{Data Availability}
The data presented in this paper can be made available upon reasonable request to the first author. We will make the light curves public and provide the breakout hydrodynamic profiles, including isotope distributions, at https://dvartany.github.io/data/.

\section*{Acknowledgments}

We thank Michael Sandoval for invaluable support with FLASH. We are also grateful to Christopher White, Anthony Piro, Matthew S.B. Coleman, and Wynn Jacobson-Galan for valuable discussion throughout the course of this project. DV acknowledges support from the NASA Hubble Fellowship Program grant HST-HF2-51520. AB acknowledges former support from the U.~S.\ Department of Energy Office of Science and the Office of Advanced Scientific Computing Research via the Scientific Discovery through Advanced Computing (SciDAC4) program and Grant DE-SC0018297 (subaward 00009650) and from the U.~S.\ National Science Foundation (NSF) under Grants AST-1714267 and PHY-1804048 (the latter via the Max-Planck/Princeton Center (MPPC) for Plasma Physics). LT acknowledges support from the Carnegie Astrophysics Summer Student Internship (CASSI) program and funding from the Rose Hills Foundation. We also acknowledge access to the Frontera cluster (under awards AST20020 and AST21003), and this research is part of the Frontera computing project at the Texas Advanced Computing Center \citep{Stanzione2020}. Frontera is made possible by NSF award OAC-1818253. Additionally, a generous award of computer time was provided by the INCITE program, enabling this research to use resources of the Argonne Leadership Computing Facility, a DOE Office of Science User Facility supported under Contract DE-AC02-06CH11357. Finally, the authors acknowledge computational resources provided by the high-performance computer center at Princeton University, which is jointly supported by the Princeton Institute for Computational Science and Engineering (PICSciE) and the Princeton University Office of Information Technology, and our continuing allocation at the National Energy Research Scientific Computing Center (NERSC), which is supported by the Office of Science of the U.~S.\ Department of Energy under contract DE-AC03-76SF00098.

\begin{deluxetable*}{lcccccc}[t]

\tablecaption{
Summary table of the progenitor, supernova, and shock breakout properties through the 17-M$_{\odot}$ simulation.
\label{tab:sn_props}}
\tablecolumns{7}
\tablewidth{0pt}
\tablehead{}
\startdata
\multicolumn{7}{c}{Progenitor Model Properties} \\
M$_{\rm ZAMS}$ & M$_{\rm cc}$ & M$_{\rm H}$ & M$_{\rm He}$  & Radius &  &  \\
(M$_{\odot}$)  & (M$_{\odot}$) & (M$_{\odot}$) & (M$_{\odot}$)  & (km) & &  \\
17 & 13.70 & 5.33 & 4.33 &   7.03$\times$10$^{8}$  \\
\hline
\multicolumn{7}{c}{Early CCSN Properties} \\
t$_{\rm run}$ & E$_{\rm exp}$ & M$_{\rm Ni}$  &  R$_{\rm shock}$ \\
 (s, pb) & (B) & (M$_{\odot}$) & (10$^4$ km) \\
 5.56  &  1.10 & 0.10 & 30$-$70\\ \hline
 \multicolumn{7}{c}{Shock Breakout Properties} \\
  t$_{\rm map}$ & t$_{\rm bo}$  & $f_{\rm mix, Ni}$ & V$_{\rm mean, Ni}$ & E$_{\rm kin}$ & E$_{\rm int}$ & t$_{\rm sim}$ \\
 (s, pb) & (day) & ($\%$) & (km\,s$^{-1}$) & (10$^{50}$\,erg) & (10$^{50}$\,erg) &(day)   \\
5.56  &  $\sim$1$-$2 & 85 & 3400 & 5.35 & 5.20 & 2$-$7\\ \hline
\hline
\enddata
\tablenotetext{}{\textbf{Note:} \textbf{Top}: Progenitor model properties. Columns are defined as follows. (1) Progenitor mass at zero-age main sequence, (2) stellar mass at core collapse, (3) mass of the hydrogen envelope at core collapse, (4) mass of the $^{4}$He envelope at core collapse, (5) stellar radius. \textbf{Middle}: Early CCSNe properties in the radiation-hydrodynamic F{\sc{ornax}} simulation. Columns are (1) CCSNe simulation duration in F{\sc{ornax}} in seconds post-bounce, (2) Explosion energy (in Bethe), (3) $^{56}$Ni mass synthesized, and shock radial extent (from minimum to maximum shock radius). \textbf{Bottom}: Explosion properties at the time of shock breakout. Columns are (1) the time of mapping to FLASH in seconds post-bounce, (2) the breakout time, including variation by direction, (3) the percent of $^{56}$Ni mass mixed outwards into the H/He interface, (4) the mean $^{56}$Ni velocity at breakout, (5) the corresponding kinetic energy, (6) the internal energy, and (7) the simulation duration in days. }
\end{deluxetable*}

\clearpage
\newpage
\bibliographystyle{aasjournal}
\bibliography{References}

\label{lastpage}
\end{document}